\documentclass[12pt]{article}
\usepackage{amssymb,epsfig}
\begin{document}
\baselineskip 24pt

\begin{titlepage}
\begin{center}
{\LARGE \bf Distribution amplitudes of $\Sigma$ and $\Lambda$ and
their electromagnetic form factors} \vspace{1cm}

{Yong-Lu Liu\footnote{E-mil: yongluliu@nudt.edu.cn} } and {Ming-Qiu
Huang}
\\[0.5cm]
\vspace*{0.1cm} {\it Department of Physics, National University of
Defense Technology, Hunan 410073, China}

\vspace{0.6cm}
\bigskip

\vskip1.2cm

{\bf Abstract \\[10pt]} \parbox[t]{\textwidth}
{\baselineskip  24pt Based on QCD conformal partial wave expansion to leading order conformal spin accuracy, we present the light-cone distribution amplitudes (DAs)
of $\Sigma$ and $\Lambda$ baryons up to twist $6$. It is concluded that fourteen independent DAs are needed to describe the valence three-quark states of the
baryons at small transverse separations. The nonperturbative parameters relevant to the DAs are determined within the framework of QCD sum rule method. With the
obtained DAs, a simple investigation on the electromagnetic form factors of these baryons are given. The magnetic moments of the baryons are estimated by fitting
the magnetic form factor with the dipole formula.}
\end{center}
\vspace{3cm}


\vspace{2mm}


\vspace{1cm}

\end{titlepage}

\newpage

\section{Introduction}
Theory of hard exclusive processes in QCD has been studied extensively for several decades. The investigation for these transitions provides unmatched opportunities
to understand the hadron structure, and the theoretical method for the calculation of these processes was developed early in 1970's \cite{exclusive,exclusive2}. In
the model of the hard exclusive process, the concept of distribution amplitudes (DAs), which are the fundamental nonperturbative functions describing the hadronic
structure, was introduced. The DAs, physically speaking, describe the decomposition of the hadron momentum in parton configurations, which is important to make the
QCD description of hard exclusive reactions quantitative.

DAs of mesons have been investigated extensively in the past, some of which have been done to high twist accuracy \cite{Braun2,Balitsky,mesondas,Agaev,Yang}.
However, the corresponding studies on baryons received less attention due to their relatively complex structure, and the existing investigations were mainly focused
on the nucleon (See \cite{review} for a review). DAs of the nucleon and other octet baryons were firstly calculated within the QCD sum rule framework on the moments
in Ref. \cite{Chernyak}. A systematic study on the nucleon DAs were provided in Ref. \cite{Braun1}, in which the DAs of the nucleon are investigated up to twist
$6$. In the paper \cite{Wang}, the axial vector higher twist DAs of the $\Lambda$ baryon were given to leading order conformal spin accuracy. Recently Ref.
\cite{Ball} gave DAs of $\Lambda_b$ in the heavy quark limit, and Ref. \cite{renom} offered a complete analysis of the one-loop renormalization of twist-$4$
operators of the nucleon. In the literature \cite{bdas}, the author presented a description on DAs of helicity $\lambda=3/2$ baryons with a new approach. Actually,
the investigation for DAs of $\Sigma$ and $\Lambda$ baryons was firstly done by Chernyak $et$ $al.$ in Ref. \cite{Chernyak}, in which they calculated the DAs up to
leading twist order only. Nevertheless, more detailed description of the internal structures of these baryons needs information on higher twist DAs. The present
work is devoted to give an investigation on $\Sigma$ and $\Lambda$ DAs up to twist $6$, and then to study their electromagnetic form factors as an application.

Higher twist contributions to DAs come from several physical origins. The first contribution is from ``bad'' components in the wave function and in particular of
components with ``wrong'' spin projection. The second one comes from transverse motion of quarks in the leading twist components. Finally, higher Fock states with
additional gluons or quark/antiquark pairs also contribute to the DAs. It has been known that for mesons, contributions due to ``bad'' components in the
quark-antiquark DAs can be described in terms of higher Fock states by equations of motion \cite{Braun2,mesondas}. Since quark-antiquark-gluon matrix elements
between the vacuum and the meson state are numerically small, contributions of ``bad'' components to mesonic DAs are small enough. However, things are different for
baryons because equations of motion are not sufficient to eliminate the higher-twist three-quark system with additional gluons. At the same time, matrix elements of
higher twist three-quark operators are large compared with the leading one. Thus the first contribution is assumed to dominate the DAs rather than the other two
origins. For this reason we only consider contributions coming from ``bad'' components in the decomposition of the Lorentz structure in this paper.

The usual description of DAs is based on the conformal symmetry of the massless QCD Lagrangian for dynamics dominated on the light-cone. DAs with definite twist can
be expanded by partial wave functions with the specific conformal spin $j$. The conformal spin of a quark is defined as $j=(l+s)/2$, where $l$ is the canonical
dimension of the quark and $s$ is its spin. For a composite particle, contributions of the higher order conformal spin $j+n$ ($n=0,1,2,$...) are given by the
leading contribution multiplied by polynomials which are orthogonal over the leading weight function. In this paper, DAs of $\Sigma$ and $\Lambda$ baryons are
investigated on the conformal partial wave expansion approach. At first glance, the mass terms of the $s$ quark break the conformal symmetry of the QCD Lagrangian
explicitly, and the $SU(3)$ breaking corrections seem difficult to be included. However, this is not a problem as argued by Ball $et$ $al.$ in the case of mesons
\cite{mesondas}. The transverse wave functions in the conformal expansion are dependent on the scale relevant to the process. If the $s$ quark mass is smaller than
the QCD scale, the transverse-momentum dependence is not affected by the quark mass. Therefore the conformal expansion of the DAs can be carried out safely. In this
work, we utilize the method proposed in Ref. \cite{Braun1} and consider the $SU(3)$ flavor symmetry breaking corrections. In fact, the $SU(3)$ flavor symmetry
breaking results in two different effects, which are isospin symmetry breaking and the corrections to the nonperturbative parameters. Unlike the case of nucleon, in
which the isospin symmetry leads to symmetric relationships to reduce the number of the independent DAs to $8$, in our case the description of $\Lambda$ and
$\Sigma$ baryons needs fourteen independent DAs that are expanded in operators with increasing conformal spin. With equations of motion, the parameters of the
conformal expansion are expressed in terms of local nonperturbative parameters, which need to be determined by nonperturbative QCD methods. In the calculation, we
expand the DAs to leading order conformal spin accuracy, and use QCD sum rules to determine the nonperturbative parameters.

DAs provide large opportunities to investigate processes connected with the baryons in the framework of light-cone sum rule (LCSR) since they are fundamental input
parameters in this framework \cite{Wang,ff,Aliev,Wym, Wzg,Huang}. LCSR is a developed nonperturbative QCD method that includes the traditional \`a la SVZ sum rule
\cite{SVZ} technique and the theory of hard exclusive processes. The main idea of LCSR is to expand the products of currents near the light-cone, and the
nonperturbative effects are described by DAs rather than condensates in the traditional QCD sum rule \cite{lcsr1,lcsr2,lcsr3}. As a simple application, the
electromagnetic (EM) form factors of the baryons are examined in LCSR with the obtained DAs. EM form factors are fundamental objects for understanding the inner
structure of the hadron. As there are no experimental data available on $\Sigma$ and $\Lambda$ EM form factors, it is instructive and necessary to give an
investigation theoretically. In the experimental point of view, the EM form factors can be described by the electric and magnetic Sachs form factors, and the
magnetic Sachs form factor at zero momentum transfer defines the magnetic moment of the baryon. It is assumed that the dependence of the magnetic form factor on the
momentum transfer can be expressed by the dipole formula, therefore after fitting the magnetic form factor by the dipole formula, we give estimations on the
magnetic moments of the $\Sigma$ baryons.

The paper is organized as follows. Section \ref{sec:def} presents notations and definitions of the baryon DAs up to twist $6$ by the matrix element of the
three-quark operator between the hadron state and the vacuum. This section also gives properties of DAs from the symmetry relationships. Section \ref{sec:conexp}
gives the leading-order conformal expansion of DAs based on the conformal invariance of the Lagrangian. The DAs are simplified to the nonperturbative parameters
which can be calculated in the framework of the QCD sum rule. Section \ref{sec:sumrule} is devoted to derive the QCD sum rules for the nonperturbative parameters
related to the DAs and then present the numerical analysis. Section \ref{sec:app} is a simple application of the obtained DAs to investigate the EM form factors of
the baryons. We give in this section the dependence of the EM form factors on the momentum transfer. After fitting the results by the dipole formula, the magnetic
moments of the baryons are estimated numerically. Summary and conclusion are given in section \ref{sec:sum}.

\section{Definitions of the light-cone distribution
amplitudes}\label{sec:def}
\subsection{General classification}

In the $SU(3)$ flavor symmetry limit, light-cone distribution amplitudes of octet baryons with quantum number $J^P=\frac12^+$ can be expressed in terms of the
matrix element of the gauge-invariant operators sandwiched between the vacuum and the baryon state:
\begin{equation}
\langle{0} |\epsilon^{ijk} {q_1}_\alpha^i(a_1 z) {q_2}_\beta^j(a_2 z) {q_3}_\gamma^k(a_3 z) |{X(P)}\rangle,\label{matr}
\end{equation}
where letters $\alpha$, $\beta$, $\gamma$ refer to Lorentz indices and $i$, $j$, $k$ refer to color indices, $q_i$ denote the quark fields, $z$ is a light-like
vector which satisfies $z^2=0$ and $a_i$ are real numbers denoting coordinates of valence quarks.

In view of the Lorentz covariance, spin and parity of the baryons, the general decomposition of the matrix element (\ref{matr}) is written as \cite{Braun1,Wang}:
\begin{eqnarray}
&& 4 \langle{0} |\epsilon^{ijk} {q_1}_\alpha^i(a_1 z) {q_2}_\beta^j(a_2 z) {q_3}_\gamma^k(a_3 z) |{X(P)}\rangle
\nonumber \\
&=& {\cal S}_1 M C_{\alpha \beta} \left(\gamma_5 X\right)_\gamma +{\cal S}_2 M^2 C_{\alpha \beta} \left(\!\not\!{z} \gamma_5 X\right)_\gamma + {\cal P}_1 M
\left(\gamma_5 C\right)_{\alpha \beta} X_\gamma + {\cal P}_2 M^2 \left(\gamma_5 C \right)_{\alpha \beta} \left(\!\not\!{z} X\right)_\gamma
\nonumber \\
&& + {\cal V}_1  \left(\!\not\!{P}C \right)_{\alpha \beta} \left(\gamma_5 X\right)_\gamma + {\cal V}_2 M \left(\!\not\!{P} C \right)_{\alpha \beta}
\left(\!\not\!{z} \gamma_5 X\right)_\gamma  + {\cal V}_3 M  \left(\gamma_\mu C \right)_{\alpha \beta}\left(\gamma^{\mu} \gamma_5 X\right)_\gamma
\nonumber \\
&& + {\cal V}_4 M^2 \left(\!\not\!{z}C \right)_{\alpha \beta} \left(\gamma_5 X\right)_\gamma + {\cal V}_5 M^2 \left(\gamma_\mu C \right)_{\alpha \beta} \left(i
\sigma^{\mu\nu} z_\nu \gamma_5 X\right)_\gamma + {\cal V}_6 M^3 \left(\!\not\!{z} C \right)_{\alpha \beta} \left(\!\not\!{z} \gamma_5 X\right)_\gamma
\nonumber \\
&& + {\cal A}_1  \left(\!\not\!{P}\gamma_5 C \right)_{\alpha \beta} X_\gamma + {\cal A}_2 M \left(\!\not\!{P}\gamma_5 C \right)_{\alpha \beta} \left(\!\not\!{z}
X\right)_\gamma  + {\cal A}_3 M \left(\gamma_\mu \gamma_5 C \right)_{\alpha \beta}\left( \gamma^{\mu} X\right)_\gamma
\nonumber \\
&& + {\cal A}_4 M^2 \left(\!\not\!{z} \gamma_5 C \right)_{\alpha \beta} X_\gamma + {\cal A}_5 M^2 \left(\gamma_\mu \gamma_5 C \right)_{\alpha \beta} \left(i
\sigma^{\mu\nu} z_\nu X\right)_\gamma + {\cal A}_6 M^3 \left(\!\not\!{z} \gamma_5 C \right)_{\alpha \beta} \left(\!\not\!{z} X\right)_\gamma
\nonumber \\
&& + {\cal T}_1 \left(P^\nu i \sigma_{\mu\nu} C\right)_{\alpha \beta} \left(\gamma^\mu\gamma_5 X\right)_\gamma + {\cal T}_2 M \left(z^\mu P^\nu i \sigma_{\mu\nu}
C\right)_{\alpha \beta} \left(\gamma_5 X\right)_\gamma \nonumber \\ &&+ {\cal T}_3 M \left(\sigma_{\mu\nu} C\right)_{\alpha \beta}
\left(\sigma^{\mu\nu}\gamma_5X\right)_\gamma+{\cal T}_4 M \left(P^\nu \sigma_{\mu\nu} C\right)_{\alpha \beta} \left(\sigma^{\mu\varrho} z_\varrho
\gamma_5X\right)_\gamma \nonumber
\\&& + {\cal T}_5 M^2 \left(z^\nu i \sigma_{\mu\nu} C\right)_{\alpha
\beta} \left(\gamma^\mu\gamma_5 X\right)_\gamma + {\cal T}_6 M^2 \left(z^\mu P^\nu i \sigma_{\mu\nu} C\right)_{\alpha \beta} \left(\!\not\!{z} \gamma_5
X\right)_\gamma \nonumber
\\ &&+ {\cal T}_7 M^2 \left(\sigma_{\mu\nu} C\right)_{\alpha \beta}
\left(\sigma^{\mu\nu} \!\not\!{z} \gamma_5 X\right)_\gamma + {\cal T}_8 M^3 \left(z^\nu \sigma_{\mu\nu} C\right)_{\alpha \beta} \left(\sigma^{\mu\varrho} z_\varrho
\gamma_5 X\right)_\gamma \,,\label{da-def}
\end{eqnarray}
where $X_\gamma$ is the spinor of the baryon, $C$ is the charge conjugation matrix and $\sigma_{\mu\nu}=\frac{i}{2}[\gamma_\mu,\gamma_\nu]$. In the present
investigations $X_\gamma$ denotes baryons with the quantum number $I(J^P)=1(\frac{1}{2}^+)$ for $\Sigma^{\pm}$ and $I(J^P)=0(\frac{1}{2}^+)$ for $\Lambda$ ($I$ is
the isospin, $J$ is the total angular momentum and $P$ is the parity). All the functions $\mathcal{S}_i$, $\mathcal {P}_i$, $\mathcal {A}_i$, $\mathcal{V}_i$ and
$\mathcal{T}_i$ depend on the scalar product $P\cdot z$.

The ``calligraphic'' invariant functions in Eq. (\ref{da-def}) do not have a definite twist however, thus the twist classification needs to be carried out in the
infinite momentum frame. Here we introduce the second auxiliary light-like vector:
\begin{equation}
p_\mu=P_\mu-\frac{1}{2}z_\mu\frac{M^2}{p\cdot z}\,, \hspace{2.5cm} p^2=0\,,
\end{equation}
so that $P\rightarrow p$ if the baryon mass can be neglected. In the infinite momentum frame, the baryon is assumed to move in the positive ${\bf e}_z$ direction,
hence $p^+$ and $z^-$ are the only nonvanishing components of $p$ and $z$. In this frame, terms on twist can be classified in powers of $p^+$, and the baryon spinor
$X_\gamma$ is decomposed into ``large'' and ``small'' components $X_\gamma^+$ and $X_\gamma^-$:
\begin{equation}
X_\gamma(P,\lambda)=\frac{1}{2p\cdot z}(\!\not\!{p}\!\not\!{z}+\!\not\!{z}\!\not\!{p})=X_\gamma^+(P,\lambda)+X_\gamma^-(P,\lambda)\,,
\end{equation}
where the projection operators
\begin{equation}
\Lambda^+=\frac{\!\not\!{p}\!\not\!{z}}{2p\cdot z}\,,\hspace{2.5cm}\Lambda^-=\frac{\!\not\!{z}\!\not\!{p}}{2p\cdot z}
\end{equation}
project the spinor onto the ``plus'' and ``minus'' components. From the Dirac equation $\!\not\! PX(P)=MX(P)$, we get the following useful relations:
\begin{equation}
\!\not\! pX(P)=MX^+(P)\,,\hspace{2.5cm}\!\not\! zX(P)=\frac{2p\cdot z}{M}X^-(P)\,.
\end{equation}
Using the explicit expressions for $X(P)$, it is easy to see that $\Lambda^+X=X^+\sim \sqrt{p^+}$ while $\Lambda^-X=X^-\sim 1/\sqrt{p^+}$. By the twist counts in
terms of $1/p^+$ the definition of light-cone DAs with a definite twist is given as
\begin{eqnarray}
&& 4\langle {0}| \epsilon^{ijk} {q_1}_\alpha^i(a_1 z) {q_2}_\beta^j(a_2 z) {q_3}_\gamma^k(a_3 z) |{X(P)} \rangle
\nonumber \\
&=& S_1 M C_{\alpha \beta} \left(\gamma_5 X^+\right)_\gamma + S_2 M C_{\alpha \beta} \left(\gamma_5 X^-\right)_\gamma + P_1 M \left(\gamma_5 C\right)_{\alpha
\beta}X^+_\gamma + P_2 M \left(\gamma_5 C \right)_{\alpha \beta} X^-_\gamma
\nonumber \\
&& + V_1  \left(\!\not\!{p}C \right)_{\alpha \beta} \left(\gamma_5 X^+\right)_\gamma + V_2  \left(\!\not\!{p}C \right)_{\alpha \beta} \left(\gamma_5
X^-\right)_\gamma + \frac{V_3}{2} M \left(\gamma_\perp C \right)_{\alpha \beta}\left( \gamma^{\perp} \gamma_5 X^+\right)_\gamma
\nonumber \\
&& + \frac{V_4}{2} M  \left(\gamma_\perp C \right)_{\alpha \beta}\left( \gamma^{\perp} \gamma_5 X^-\right)_\gamma + V_5 \frac{M^2}{2 p z} \left(\!\not\!{z}C
\right)_{\alpha \beta} \left(\gamma_5 X^+\right)_\gamma + \frac{M^2}{2 pz} V_6 \left(\!\not\!{z}C \right)_{\alpha \beta} \left(\gamma_5 X^-\right)_\gamma
\nonumber \\
&& + A_1  \left(\!\not\!{p}\gamma_5 C \right)_{\alpha \beta} X^+_\gamma + A_2 \left(\!\not\!{p}\gamma_5 C \right)_{\alpha \beta} X^-_\gamma + \frac{A_3}{2} M
\left(\gamma_\perp \gamma_5 C \right)_{\alpha \beta}\left( \gamma^{\perp} X^+\right)_\gamma
\nonumber \\
&& + \frac{A_4}{2} M  \left(\gamma_\perp \gamma_5 C \right)_{\alpha \beta}\left( \gamma^{\perp} X^-\right)_\gamma + A_5  \frac{M^2}{2 p z} \left(\!\not\!{z}\gamma_5
C \right)_{\alpha \beta} X^+_\gamma + \frac{M^2}{2 pz}  A_6 \left(\!\not\!{z}\gamma_5 C \right)_{\alpha \beta} X^-_\gamma
\nonumber \\
&& + T_1 \left(i \sigma_{\perp p} C\right)_{\alpha \beta} \left(\gamma^\perp\gamma_5 X^+\right)_\gamma + T_2 \left(i \sigma_{\perp\, p} C\right)_{\alpha \beta}
\left(\gamma^\perp\gamma_5 X^-\right)_\gamma + T_3 \frac{M}{p z} \left(i \sigma_{p\, z} C\right)_{\alpha \beta} \left(\gamma_5 X^+\right)_\gamma
\nonumber \\
&& + T_4 \frac{M}{p z}\left(i \sigma_{z\, p} C\right)_{\alpha \beta} \left(\gamma_5 X^-\right)_\gamma + T_5 \frac{M^2}{2 p z}  \left(i \sigma_{\perp\, z}
C\right)_{\alpha \beta} \left(\gamma^\perp\gamma_5 X^+\right)_\gamma
\nonumber \\
&&+ \frac{M^2}{2 pz}  T_6 \left(i \sigma_{\perp\, z} C\right)_{\alpha \beta} \left(\gamma^\perp\gamma_5 X^-\right)_\gamma + M \frac{T_7}{2} \left(\sigma_{\perp\,
\perp'} C\right)_{\alpha \beta} \left(\sigma^{\perp\, \perp'} \gamma_5 X^+\right)_\gamma \nonumber\\ &&+ M \frac{T_8}{2} \left(\sigma_{\perp\, \perp'}
C\right)_{\alpha \beta} \left(\sigma^{\perp\, \perp'} \gamma_5 X^-\right)_\gamma\,,\label{da-deftwist}
\end{eqnarray}
where an obvious notation $\sigma_{pz}=\sigma^{\mu\nu}p_\mu z_\nu$, etc., is used as a shorthand and $\perp$ stands for the projection transverse to $z,p$, e.g.
$\gamma_\perp\gamma^\perp=\gamma^\mu g_{\mu\nu}^\perp\gamma^\nu$ with $g_{\mu\nu}^\perp=g_{\mu\nu}-(p_\mu z_\nu+z_\mu p_\nu)/pz$. The DAs $F=S_i,P_i,V_i,A_i,T_i$
with a definite twist are classified in Table \ref{tabDA-def}. Each distribution amplitude $F_i$ can be represented as
\begin{equation}
F(a_ip\cdot z)=\int \mathcal Dxe^{-ipz\sum_ix_ia_i}F(x_i)\,,
\end{equation}
where the dimensionless variables $x_i$, which satisfy the relation $0<x_i<1,\; \sum_ix_i=1$, correspond to the longitudinal momentum fractions carried by the
quarks inside the baryon. The integration measure is defined as
\begin{equation}
\int \mathcal Dx=\int_0^1dx_1dx_2dx_3\delta(x_1+x_2+x_3-1)\,.
\end{equation}
Comparing Eq. (\ref{da-def}) and (\ref{da-deftwist}), the invariant functions $\mathcal S_i,\mathcal P_i,\mathcal V_i,\mathcal A_i,\mathcal T_i$ can be expressed in
terms of the DAs $S_i,P_i,V_i,A_i,T_i$ with a definite twist.

For scalar and pseudo-scalar distributions the following relations hold:
\begin{eqnarray}
\renewcommand{\arraystretch}{1.7}
\hspace*{-4.7cm}
\begin{array}{lll}
 {\cal S}_1 = S_1\,, &\qquad\qquad\qquad& 2p\cdot z\, {\cal S}_2 = S_1-S_2\,, \\
 {\cal P}_1 = P_1\,, &\qquad\qquad\qquad& 2p\cdot z\, {\cal P}_2 = P_2-P_1\,,
\end{array}
\renewcommand{\arraystretch}{1.0}
\end{eqnarray}
for vector distributions:
 \begin{eqnarray}
\renewcommand{\arraystretch}{1.7}
\begin{array}{lll}
 \mathcal V_1 = V_1\,, &~~& 2 p\cdot z \mathcal V_2 = V_1 - V_2 - V_3\,, \\
 2 \mathcal V_3 = V_3\,, &~~& 4 p\cdot z \mathcal V_4 = - 2 V_1 + V_3 + V_4  + 2 V_5\,, \\
4 p\cdot z \mathcal V_5 = V_4 - V_3\,, &~~& (2 p\cdot z )^2 \mathcal
V_6 = - V_1 + V_2 +  V_3 +  V_4 + V_5 - V_6\,,
\end{array}
\renewcommand{\arraystretch}{1.0}
\end{eqnarray}
 for axial vector distributions:
 \begin{eqnarray}
\renewcommand{\arraystretch}{1.7}
\begin{array}{lll}
 \mathcal A_1 = A_1\,, &~~& 2 p\cdot z \mathcal A_2 = - A_1 + A_2 -  A_3\,, \\
 2 \mathcal A_3 = A_3\,, &~~& 4 p\cdot z \mathcal A_4 = - 2 A_1 - A_3 - A_4  + 2 A_5\,, \\
 4 p\cdot z \mathcal A_5 = A_3 - A_4\,, &~~&
(2 p\cdot z )^2  \mathcal A_6 =  A_1 - A_2 +  A_3 +  A_4 - A_5 +
A_6\,,
\end{array}
\renewcommand{\arraystretch}{1.0}
\end{eqnarray}
 and, finally, for tensor distributions:
\begin{eqnarray}
\renewcommand{\arraystretch}{1.7}
\begin{array}{lll}
 \mathcal T_1 = T_1\,, &~~& 2 p\cdot z \mathcal T_2 = T_1 + T_2 - 2 T_3\,, \\
2 \mathcal T_3 = T_7\,, &~~& 2 p\cdot z \mathcal T_4 = T_1 - T_2 - 2  T_7\,, \\
2 p\cdot z \mathcal T_5 = - T_1 + T_5 + 2  T_8\,, &~~&
    (2 p\cdot z)^2 \mathcal T_6 = 2 T_2 - 2 T_3 - 2 T_4 + 2 T_5 + 2 T_7 + 2 T_8\,,\\
4 p\cdot z \mathcal T_7 = T_7 - T_8\,, &~~& (2 p\cdot z)^2 \mathcal
T_8 = -T_1 + T_2 + T_5 - T_6 + 2 T_7 + 2 T_8 \,.
\end{array}
\renewcommand{\arraystretch}{1.0}
\end{eqnarray}

For $\Sigma^{+(-)}$, the identity of the two $u(d)$ quarks in the baryon gives symmetry properties of the DAs. The Lorentz decomposition on the $\gamma$-matrix
structure implies that the vector and tensor DAs are symmetric, whereas the scalar, pseudo-scalar and axial-vector DAs are antisymmetric under the interchange of
the two $u(d)$ quarks:
\begin{eqnarray}
V_i(1,2,3)&=&\;\ V_i(2,1,3)\,,\hspace{2.8cm} T_i(1,2,3)=\;\ T_i(2,1,3)\,,\nonumber\\
S_i(1,2,3)&=&-S_i(2,1,3)\,,\hspace{2.7cm} P_i(1,2,3)=-P(2,1,3)\,,\nonumber\\
A_i(1,2,3)&=&-A(2,1,3)\,.
\end{eqnarray}
The ``calligraphic'' structures in Eq. (\ref{da-def}) have the similar relationships.

For $\Lambda$,  the isospin symmetry leads to similar relationships that the vector and tensor DAs are antisymmetric, while the scalar, pseudoscalar and
axial-vector DAs are symmetric:
\begin{eqnarray}
V_i(1,2,3)&=&-V_i(2,1,3)\,,\hspace{2.8cm} T_i(1,2,3)=-T_i(2,1,3)\,,\nonumber\\
S_i(1,2,3)&=&\;\ S_i(2,1,3)\,,\hspace{2.8cm} P_i(1,2,3)=\;\ P(2,1,3)\,,\nonumber\\
A_i(1,2,3)&=&\;\ A(2,1,3)\,.
\end{eqnarray}
\subsection{Representation in terms of chiral fields}
This subsection gives the DAs' representation in terms of chiral fields. The discussion is mainly about $\Sigma^+$ baryon, and the counterparts of the others are
similar. In terms of quark fields with definite chirality:
\begin{equation}
q^{\uparrow({\downarrow})}=\frac{1}{2}(1\pm\gamma_5)q\,,
\end{equation}
the DAs can be interpreted transparently. Projection on the state where the spins of the two $u$ quarks are antiparallel, that is $u^\uparrow u^\downarrow$, singles
out vector and axial vector amplitudes, while the two $u$ quarks are parallel, $u^\uparrow u^\uparrow$ and $u^\downarrow u^\downarrow$, singles out scalar,
pseudo-scalar and tensor structures. Similar as expressions in Ref. \cite{Braun1}, the DAs can be defined in terms of chiral fields. The leading twist-$3$
distribution amplitude can be defined as:
\begin{eqnarray}
\langle{0}| \epsilon^{ijk}\! \left(u^{\uparrow}_i(a_1 z) C \!\!\not\!{z} u^{\downarrow}_j(a_2 z)\right) \!\not\!{z} s^{\uparrow}_k(a_3 z)| {P}\rangle &=& - \frac12
pz \!\not\!{z} {\Sigma^+}^\uparrow \int \mathcal D x \,e^{-i pz \sum x_i a_i}\, \Phi_3(x_i),\nonumber\\ \label{def-chi1}
\end{eqnarray}
and the distributions for twist-$4$ are:
\begin{eqnarray}
\langle{0}| \epsilon^{ijk}\! \left(u^{\uparrow}_i(a_1 z) C \!\!\not\!{z} u^{\downarrow}_j(a_2 z)\right) \!\not\!{p} s^{\uparrow}_k(a_3 z)| {P}\rangle &=& - \frac12
pz \!\not\!{p} {\Sigma^+}^\uparrow\!\! \int\! \mathcal D x \,e^{-i pz \sum x_i a_i}\, \Phi_4(x_i),
\nonumber \\
\langle{0}| \epsilon^{ijk}\! \left(u^{\uparrow}_i(a_1 z) C\!\not\!{z}\gamma_\perp\!\!\not\!{p}
 u^{\downarrow}_j(a_2 z)\right)
\gamma^\perp\! \!\!\not\!{z} s^{\downarrow}_k(a_3 z) |{P}\rangle &=& - pz M \!\!\not\!{z} {\Sigma^+}^\uparrow\! \int\!\! \mathcal D x \,e^{-i pz \sum x_i a_i}\,
\Psi_4(x_i),
 \nonumber\\
\langle{0}| \epsilon^{ijk}\! \left(u^{\uparrow}_i(a_1 z) C \!\not\!{p}\!\!\not\!{z} u^{\uparrow}_j(a_2 z)\right) \!\not\!{z} s^{\uparrow}_k(a_3 z)|{P}\rangle &=&
\frac12 pz M \!\not\!{z} {\Sigma^+}^\uparrow\! \int\!\! \mathcal D x \,e^{-i pz \sum x_i a_i}\, \Xi_4(x_i),\nonumber\\ \label{def-chi2}
\end{eqnarray}
and the distributions for twist-$5$ are similarly written as:
\begin{eqnarray}
\langle{0}| \epsilon^{ijk}\! \left(u^{\uparrow}_i(a_1 z) C \!\not\!{p} u^{\downarrow}_j(a_2 z)\right) \!\not\!{z} s^{\uparrow}_k(a_3 z)|{P} \rangle &=& - \frac14
M^2 \!\not\!{z} {\Sigma^+}^\uparrow\! \int\!\! \mathcal D x \,e^{-i pz \sum x_i a_i}\, \Phi_5(x_i),
\nonumber \\
\langle{0}| \epsilon^{ijk}\! \left(u^{\uparrow}_i(a_1 z) C \!\!\not\!{p}\gamma_\perp\!\!\not\!{z}
 u^{\downarrow}_j(a_2 z)\right)
\gamma^\perp \!\!\not\!{p}\, s^{\downarrow}_k(a_3 z)|{P} \rangle &=& - pz M \!\not\!{p}\, {\Sigma^+}^\uparrow \int \mathcal D x \,e^{-i pz \sum x_i a_i}\,
\Psi_5(x_i),
\nonumber \\
\langle{0}| \epsilon^{ijk}\! \left(u^{\uparrow}_i(a_1 z) C\!\!\not\!{z}\!\!\not\!{p}\, u^{\uparrow}_j(a_2 z)\right) \!\not\!{p} s^{\uparrow}_k(a_3 z)|{P} \rangle
&=&  \frac12 pz M \!\not\!{p}\, {\Sigma^+}^\uparrow\! \int\!\! \mathcal D x \,e^{-i pz \sum x_i a_i}\, \Xi_5(x_i),\nonumber \\ \label{def-chi3}
\end{eqnarray}
and finally the twist-$6$ one can be expressed as:
\begin{eqnarray}
\langle{0}| \epsilon^{ijk}\! \left(u^{\uparrow}_i(a_1 z) C \!\!\not\!{p} u^{\downarrow}_j(a_2 z)\right) \!\not\!{p} s^{\uparrow}_k(a_3 z)|{P} \rangle &=& - \frac14
M^2 \!\not\!{p} {\Sigma^+}^\uparrow\! \int\!\! \mathcal D x \,e^{-i pz \sum x_i
a_i}\, \Phi_6(x_i).\nonumber\\
\label{def-chi4}
\end{eqnarray}

Due to $SU(3)$ flavor symmetry breaking effects, there are not similar relationships as that in Ref. \cite{Braun1} from the isospin symmetry to reduce the number of
the independent DAs. So the following additional chiral fields representations are needed to get all the DAs:
\begin{eqnarray}
\langle{0}| \epsilon^{ijk}\! \left(u^{\downarrow}_i(a_1 z) C \!\not\!{p}\!\!\not\!{z} u^{\downarrow}_j(a_2 z)\right) \!\not\!{z} s^{\uparrow}_k(a_3 z)|{P}\rangle
&=&  \frac12 pz M \!\not\!{z} {\Sigma^+}^\uparrow\! \int\!\! \mathcal D x \,e^{-i pz \sum x_i
a_i}(S_1-P_1+T_3+T_7), \nonumber\\
\langle{0}| \epsilon^{ijk}\! \left(u^{\downarrow}_i(a_1 z) C\!\!\not\!{z}\!\!\not\!{p}\, u^{\downarrow}_j(a_2 z)\right) \!\not\!{p} s^{\uparrow}_k(a_3 z)|{P}
\rangle &=&  \frac12 pz M \!\not\!{p}\, {\Sigma^+}^\uparrow\! \int\!\! \mathcal D x \,e^{-i pz \sum x_i a_i}(S_2-P_2-T_4+T_8),\nonumber\\ \label{def-chi5}
\end{eqnarray}
and
\begin{eqnarray}
\langle{0}| \epsilon^{ijk}\! \left(u^{\uparrow}_i(a_1 z) C i \sigma_{\perp z} u^{\uparrow}_j(a_2 z)\right) \gamma^\perp\!\not\!{z} s^{\downarrow}_k(a_3
z)|{P}\rangle &=&  - 2 pz \!\not\!{z} {\Sigma^+}^\uparrow\! \!\int\!\! \mathcal D x \,e^{-i pz \sum x_i a_i}\, T_1(x_i),
\nonumber \\
\langle{0}| \epsilon^{ijk}\! \left(u^{\downarrow}_i(a_1 z) C i \sigma_{\perp z} u^{\downarrow}_j(a_2 z)\right) \gamma^\perp\!\not\!{p} s^{\downarrow}_k(a_3
z)|{P}\rangle &=&  - 2 pz \!\not\!{p} {\Sigma^+}^\uparrow\! \!\int\!\! \mathcal D x \,e^{-i pz \sum x_i a_i}\, T_2(x_i),
\nonumber \\
\langle{0}| \epsilon^{ijk}\! \left(u^{\downarrow}_i(a_1 z) C i \sigma_{\perp p} u^{\downarrow}_j(a_2 z)\right) \gamma^\perp\!\not\!{z} s^{\downarrow}_k(a_3 z)|{P}
\rangle &=&  - M^2 \!\not\!{z} {\Sigma^+}^\uparrow\! \!\int\!\! \mathcal D x \,e^{-i pz \sum x_i a_i}\, T_5(x_i),
\nonumber \\
\langle{0}| \epsilon^{ijk}\! \left(u^{\uparrow}_i(a_1 z) C i \sigma_{\perp p} u^{\uparrow}_j(a_2 z)\right) \gamma^\perp\!\not\!{p} s^{\downarrow}_k(a_3 z)|{P}
\rangle &=&  - M^2\!\not\!{p}{\Sigma^+}^\uparrow\! \!\int\!\! \mathcal D x \,e^{-i
pz \sum x_i a_i}\, T_6(x_i).\nonumber\\
\label{def-chi6}
\end{eqnarray}
The twist classification of these additional DAs are shown in Table \ref{tabDA-def}. The following denotations $(S_1-P_1+T_3+T_7)(x_i)\equiv \Xi_4'(x_i)$, and
$(S_2-P_2-T_4+T_8)(x_i)\equiv \Xi_5'(x_i)$ are adopted for convenience.

The similar relationships hold for the $\Lambda$ baryon under the exchange $u,u,s$ to $u,d,s$, and for the $\Sigma^-$ baryon under the exchange $u,u,s$ to $d,d,s$.
\section{Conformal expansion}\label{sec:conexp}
The spirit of the conformal expansion of distribution amplitudes is similar to the partial wave expansion of a wave function in  quantum mechanics. The idea is to
use the conformal symmetry of the massless QCD Lagrangian to study the DAs, which allows to separate longitudinal degrees of freedom from transverse ones
\cite{Braun2,Yang,Braun1}. The transverse coordinates are replaced by the renormalization scale, which is determined by the renormalization group. The dependence on
longitudinal momentum fractions, which is living on the light cone, is taken into account by a set of orthogonal polynomials that form an irreducible representation
of the collinear subgroup $SL(2,R)$ of the conformal group. As to leading logarithmic accuracy, the renormalization group equations are driven by tree-level counter
terms, they have the conformal symmetry. This leads to the fact that components of the DAs with different conformal spin do not mix under renormalization to this
accuracy.

The $SL(2,R)$ group is governed by four generators ${\bf P}_+$, ${\bf M}_{-+}$, ${\bf D}$ and ${\bf K}_-$, where the definitions are used for a vector $A$:
$A_+=A_\mu z^\mu$ and $A_-=A_\mu p^\mu/p\cdot z$. The four generators ${\bf P}_\mu$, ${\bf K}_\mu$, $\bf D$ and ${\bf M}_{\mu\nu}$ are the translation, special
conformal transformation, dilation and Lorentz generators, respectively. The generators of the collinear subgroup $SL(2,R)$ can be described by the following four
operators:
\begin{equation}
{\bf L}_+=-i{\bf P}_+\,,\; {\bf L}_-=\frac{i}{2}{\bf K}_-\,,\;  {\bf L}_0=-\frac{i}{2}({\bf D}-{\bf M}_{-+})\,,\;  {\bf E}=i({\bf D}+{\bf M}_{-+})\,.
\end{equation}
For a field living on the light cone $\Phi(z)$, the acting of the above generators on it yields the following relations:
\begin{eqnarray}
&&[{\bf L}^2,\Phi(z)]=j(j-1)\Phi(z)\,,\hspace{2.0cm} [{\bf E},\Phi(z)]=(l-s)\Phi(z)\,,\nonumber\\
\;\ &&[{\bf E},{\bf L}^2]=0\,, \hspace{4.6cm} [{\bf E},{\bf L}_0]=0\,,
\end{eqnarray}
and
\begin{equation}
{\bf L}^2={\bf L}_0^2-{\bf L}_0+{\bf L}_+{\bf L}_-\,,
\end{equation}
where $j=(l+s)/2$ is called the conformal spin and $t=l-s$ is the twist. In above notations, $l$ is the canonical dimension of the quark field and $s$ is the quark
spin projection on the light-cone. The role of the generator ${\bf E}$ is analogous to the Hamiltonian in quantum mechanics, and the twist corresponds to the
eigenvalue of the Hamiltonian. For a given twist distribution amplitude, it can be expanded by the conformal partial wave functions that are the eigenstates of
${\bf L}^2$ and $L_0$.

For multi-quark states, we need to deal with the problem of summation of conformal spins, and here the group is non-compact. The distribution amplitude with the
lowest conformal spin $j_{min}=j_1+j_2+j_3$ of a three-quark state is \cite{Braun2,Balitsky}
\begin{equation}
\Phi_{as}(x_1,x_2,x_3)=\frac{\Gamma[2j_1+2j_2+2j_3]}{\Gamma[2j_1]\Gamma[2j_2]\Gamma[2j_3]}{x_1}^{2j_1-1}{x_2}^{2j_2-1}{x_3}^{2j_3-1}\,.\label{as-dis}
\end{equation}
Contributions with higher conformal spin $j=j_{min}+n$ ($n=1,2,...$) are given by $\Phi_{as}$ multiplied by polynomials that are orthogonal over the weight function
(\ref{as-dis}). In this paper, we just consider DAs to leading order conformal spin accuracy. For DAs in Table \ref{tabDA-def}, we give their conformal expansion:
\begin{eqnarray}
\Phi_3(x_i)=120x_1x_2x_3\phi_3^0(\mu),\hspace{2.5cm} T_1(x_i)=120x_1x_2x_3\phi_3'^0(\mu)\,,\label{contwist3}
\end{eqnarray}
for twist $3$ and
\begin{eqnarray}
\Phi_4(x_i)&=&24x_1x_2\phi_4^0(\mu)\,,\hspace{2.5cm}
\Psi_4(x_i)=24x_1x_3\psi_4^0(\mu)\,,\nonumber\\
\Xi_4(x_i)&=&24x_2x_3\xi_4^0(\mu)\,,\hspace{2.5cm}
\Xi_4'(x_i)=24x_2x_3\xi_4'^0(\mu)\,,\nonumber\\
T_2(x_i)&=&24x_1x_2\phi_4'(\mu)\,,\label{contwist4}
\end{eqnarray}
for twist $4$ and
\begin{eqnarray}
\Phi_5(x_i)&=&6x_3\phi_5^0(\mu)\,,\hspace{2.5cm}
\Psi_5(x_i)=6x_2\psi_5^0(\mu)\,,\nonumber\\
\Xi_5(x_i)&=&6x_1\xi_5^0(\mu)\,,\hspace{2.5cm}
\Xi_5'(x_i)=6x_1\xi_5'^0(\mu)\,,\nonumber\\
T_5(x_i)&=&6x_3\phi_5'(\mu)\,,\label{contwist5}
\end{eqnarray}
for twist $5$ and
\begin{eqnarray}
\Phi_6(x_i)=2\phi_6^0(\mu)\,,\hspace{3.5cm} T_6(x_i)=2\phi_6'(\mu)\,.\label{contwist6}
\end{eqnarray}
for twist $6$. There are altogether $14$ parameters which can be determined by the equations of motion.
\subsection{DAs of the $\Sigma$ baryon}
The normalization of the DAs of $\Sigma^+$ are determined by matrix element of the local three-quark operator. The Lorentz decomposition of the matrix element can
be expressed explicitly as follows:
\begin{eqnarray}
4\langle0|\epsilon^{ijk}u^i_\alpha(0)u^j_\beta(0)s^k_\gamma(0)|\Sigma^+(P)\rangle=\mathcal{V}^0_1(\!\not\!
PC)_{\alpha\beta}(\gamma_5\Sigma^+)_\gamma+\mathcal{V}^0_3(\gamma_\mu
C)_{\alpha\beta}(\gamma_\mu\gamma_5\Sigma^+)_\gamma\nonumber\\
+\mathcal{T}^0_1(P^\nu i\sigma_{\mu\nu}C)_{\alpha \beta}(\gamma^\mu\gamma_5\Sigma^+)_\gamma+\mathcal{T}^0_3M(\sigma_{\mu\nu}C)_{\alpha
\beta}(\sigma^{\mu\nu}\gamma_5\Sigma^+)_\gamma\,.
\end{eqnarray}
Similar to definitions in Ref. \cite{Braun1}, the above four parameters can be expressed by the following matrix elements:
\begin{eqnarray}
&& \langle{0}| \epsilon^{ijk} \left[u^i(0) C \!\not\!{z} u^j(0)\right] \, \gamma_5 \!\not\!{z} s^k(0)| {P}\rangle =   f_{\rm \Sigma^+}
(p\cdot z) \!\not\!{z} \Sigma^+(P)\,,  \nonumber \\
&&\langle{0}| \epsilon^{ijk} \left[u^i(0) C\gamma_\mu u^j(0)\right]\, \gamma_5 \gamma^\mu s^k(0)| {P}\rangle = \lambda_1 M \Sigma^+(P) \,,
\nonumber \\
&&\langle{0}| \epsilon^{ijk} \left[u^i(0) C\sigma_{\mu\nu} u^j(0)\right] \, \gamma_5 \sigma^{\mu\nu} s^k(0)|{P}\rangle = \lambda_2 M \Sigma^+(P) \,,
\nonumber\\
&&\langle{0}| \epsilon^{ijk} \left[u^i(0) Ciq^\nu\sigma_{\mu\nu}u^j(0)\right]\,\gamma_5\gamma_\mu s^k(0)|{P}\rangle=\lambda_3 M\!\not\!{q}\Sigma^+(P)
\,.\label{def-nonlocal}
\end{eqnarray}
As mentioned above, there are no relations derived from isospin symmetry, so four but not three matrix elements are needed to determine the four parameters
$\mathcal V_1^0, \mathcal V_3^0, \mathcal T_1^0$, and $\mathcal T_3^0$. After a simple calculation, we arrive at the following expressions of $\mathcal V_1^0,
\mathcal V_3^0, \mathcal T_1^0$, and $\mathcal T_3^0$ with the four parameters defined in Eqs. (\ref{def-nonlocal}):
\begin{eqnarray}
\mathcal V_1^0&=&f_{\Sigma^+}\,,\hspace{3.8cm}\mathcal V_3^0=\frac
14(f_{\Sigma^+}-\lambda_1)\,,\nonumber\\
\mathcal T_1^0&=&\frac 16(4\lambda_3-\lambda_2)\,,\hspace{2.1cm}\mathcal T_3^0=\frac 1{12}(2\lambda_3-\lambda_2)\,.\label{nonlocalpara}
\end{eqnarray}

At the same time, coefficients of operators in Eqs. (\ref{contwist3})-(\ref{contwist6}) can be expressed  to leading order conformal spin accuracy as
\begin{eqnarray}
\phi_3^0&=&\phi_6^0=f_{\Sigma^+},\hspace{2.8cm}\psi_4^0=\psi_5^0=\frac12(f_{\Sigma^+}-\lambda_1)\,,\nonumber\\
\phi_4^0&=&\phi_5^0=\frac12(f_{\Sigma^+}+\lambda_1),\hspace{1.3cm}\phi_3'^0=\phi_6'^0=-\xi_5^0=\frac16(4\lambda_3-\lambda_2)\,,\nonumber\\
\phi_4'^0&=&\xi_4^0=\frac16(8\lambda_3-3\lambda_2),\hspace{1.2cm}\phi_5'^0=-\xi_5'^0=\frac16\lambda_2\,,\nonumber\\
\xi_4'^0&=&\frac16(12\lambda_3-5\lambda_2)\,.
\end{eqnarray}
\subsection{DAs of the $\Lambda$ baryon}
The Lorentz decomposition of the local matrix element of $\Lambda$ can be expressed explicitly as follows:
\begin{eqnarray}
4\langle0|\epsilon^{ijk}u^i_\alpha(0)d^j_\beta(0)s^k_\gamma(0)|\Lambda(P)\rangle=\mathcal{S}^0_1C_{\alpha\beta}(\gamma_5\Lambda)_\gamma+\mathcal{P}^0_1(\gamma_5
C)_{\alpha\beta}\Lambda_\gamma\nonumber\\
+\mathcal{A}^0_1(\!\not\!P\gamma_5C)_{\alpha \beta}\Lambda_\gamma+\mathcal{A}^0_3M(\gamma_\mu\gamma_5C)_{\alpha \beta}(\gamma^\mu\Lambda)_\gamma\,.
\end{eqnarray}
In order to get the above parameters, the following matrix elements are used:
\begin{eqnarray}
&& \langle{0}| \epsilon^{ijk} \left[u^i(0) C\gamma_5 \!\not\!{z} d^j(0)\right] \!\not\!{z} s^k(0)| {P}\rangle = f_{\Lambda}
(p\cdot z) \!\not\!{z} \Lambda(P)\,,  \nonumber \\
&&\langle{0}| \epsilon^{ijk} \left[u^i(0) C\gamma_5\gamma_\mu d^j(0)\right]\, \gamma^\mu s^k(0)| {P}\rangle = \lambda_1 M \Lambda(P) \,,
\nonumber \\
&&\langle{0}| \epsilon^{ijk} \left[u^i(0) C\gamma_5 d^j(0)\right] \, s^k(0)|{P}\rangle = \lambda_2 M \Lambda(P) \,,
\nonumber\\
&&\langle{0}| \epsilon^{ijk} \left[u^i(0) Cd^j(0)\right]\,\gamma_5 s^k(0)|{P}\rangle=\lambda_3 M^2\Lambda(P) \,.
\end{eqnarray}
A simple calculation leads to the following relationships:
\begin{eqnarray}
\mathcal A_1^0&=&f_\Lambda,\hspace{2.8cm}\mathcal A_3^0=-\frac
14(f_\Lambda-\lambda_1),\nonumber\\
\mathcal P_1^0&=&\lambda_2,\hspace{2.8cm}\mathcal S_1^0=\lambda_3.
\end{eqnarray}

To leading order of the conformal spin expansion, coefficients of operators in Eqs. (\ref{contwist3})-(\ref{contwist6}) for $\Lambda$ can be expressed as
\begin{eqnarray}
\phi_3^0&=&\phi_6^0=-f_\Lambda,\hspace{2.8cm}\phi_4^0=\phi_5^0=-\frac12(f_\Lambda+\lambda_1),\nonumber\\
\psi_4^0&=&\psi_5^0=\frac12(f_\Lambda-\lambda_1),\hspace{1.6cm}\xi_4^0=\xi_5^0=\lambda_2+\lambda_3,\nonumber\\
\xi_4^{'0}&=&\xi_5^{'0}=\lambda_3-\lambda_2.
\end{eqnarray}
\section{Determination of the parameters in QCD sum rules}\label{sec:sumrule}
\subsection{QCD sum rules for $\Sigma$ baryon}
Determination of the nonperturbative parameters $f_{\Sigma}$, $\lambda_1$, $\lambda_2$ and $\lambda_3$ can be done in the framework of QCD sum rule. The method is
carried out from the following correlation functions for $\Sigma^+$:
\begin{equation}
\Pi_i(q^2)=\int d^4x e^{iq\cdot x}\langle 0|T\{j_i(x)\bar j_i(0)\}|0\rangle,
\end{equation}
with the definitions of the currents:
\begin{eqnarray}
j_1(x)&=&\epsilon^{ijk}[u^i(x)C\!\not\! {z}u^j(x)]\gamma_5\!\not\! {z}s^k(x),\label{CZcurrent}\\
j_2(x)&=&\epsilon^{ijk}[u^i(x)C\gamma_\mu u^j(x)]\gamma_5\gamma^\mu s^k(x),\\
j_3(x)&=&\epsilon^{ijk}[u^i(x)C\sigma_{\mu\nu} u^j(x)]\gamma_5\sigma^{\mu\nu}s^k(x),
\end{eqnarray}
and the forth current:
\begin{equation}
j_4(x)=\epsilon^{ijk}[u^i(x)Ciq^\nu\sigma_{\mu\nu} u^j(x)]\gamma_5\gamma^\mu s^k(x).
\end{equation}
Inserting the complete set of states with the same quantum numbers as those of $\Sigma^+$, the hadronic representations of the correlation functions are given as
\begin{eqnarray}
\Pi_1(q^2)&=&2f_{\Sigma^+}^2(q\cdot z)^3\!\not\!
{z}\frac{1}{M^2-q^2}+\int_{s_0}^\infty\frac{\rho_1^h(s)}{s-q^2}ds,\nonumber\\
\Pi_2(q^2)&=&M^2\lambda_1^2\frac{\!\not\!q+M}{M^2-q^2}+\int_{s_0}^\infty\frac{\rho_2^h(s)}{s-q^2}ds,\nonumber\\
\Pi_3(q^2)&=&M^2\lambda_2^2\frac{\!\not\!q+M}{M^2-q^2}+\int_{s_0}^\infty\frac{\rho_3^h(s)}{s-q^2}ds,\nonumber\\
\Pi_4(q^2)&=&q^2M^2\lambda_3^2\frac{\!\not\!q+M}{M^2-q^2}+\int_{s_0}^\infty\frac{\rho_4^h(s)}{s-q^2}ds.
\end{eqnarray}
On the operator product expansion (OPE) side, condensates up to dimension $6$ are taken into account. To give the sum rules, we utilize the dispersion relationship
and assume the quark-hadron duality. After taking Borel transformation on both sides of the hadronic representation and QCD expansion, and equating the two sides,
the final sum rules are given as follows:
\begin{equation}
4(2\pi)^4f_{\Sigma^+}^2e^{-\frac{M^2}{M_B^2}}=\frac15\int_{m_s^2}^{s_0}s(1-x)^5e^{-\frac{s}{M_B^2}}ds-\frac{b}{6}
\int_{m_s^2}^{s_0}\frac{x(1-x)(1-2x)}{s}e^{-\frac{s}{M_B^2}}ds,
\end{equation}
and
\begin{eqnarray}
4(2\pi)^4\lambda_1^2M^2e^{-\frac{M^2}{M_B^2}}&=&\int_{m_s^2}^{s_0}s^2[(1-x)(1+x)(1-8x+x^2)-12x^2\ln x]e^{-\frac{s}{M_B^2}}ds\nonumber\\
&&+\frac{b}{6}\int_{m_s^2}^{s_0}(1-x)^2e^{-\frac{s}{M_B^2}}ds+\frac83a^2(1-\frac{m_0^2}{2M_B^2}-\frac{m_0^2m_s^2}{2M_B^4}\nonumber\\
&& +\frac{m_0^4m_s^4}{16M_B^8})e^{-\frac{m^2}{M_B^2}}-2a_sm_s\int_{m_s^2}^{s_0}e^{-\frac{s}{M_B^2}}ds,
\end{eqnarray}
and
\begin{eqnarray}
2(2\pi)^4\lambda_2^2M^2e^{-M^2/M_B^2}=-\int_{m_s^2}^{s_0}s^2(-1+8x-8x^3+x^4+12x^2\ln x)e^{-\frac{s}{M_B^2}}ds\nonumber\\
+\frac b3\int_{m_s^2}^{s_0}(1-x)(4-7x)e^{-\frac{s}{M_B^2}}ds+12m_sa_s\int_{m_s^2}^{s_0}e^{-\frac{s}{M_B^2}}ds,
\end{eqnarray}
and
\begin{eqnarray}
(4\pi)^4\lambda_3^2M^2e^{-M^2/M_B^2}&=&\int_{m_s^2}^{s_0}s^2\{[(1-x)(1+x)(1-8x+x^2)-12x^2\ln x]\nonumber\\
&&+\frac15(1-x)^5\}e^{-\frac{s}{{M_B}^2}}ds+\frac{b}{12}
\int_{m_s^2}^{s_0}(1-x)(11-5x-4x^2)e^{-\frac{s}{{M_B}^2}}ds\nonumber\\
&&+16a^2(1-\frac{m_0^2}{2M_B^2}-\frac{m_0^2m_s^2}{2M_B^4}
+\frac{m_0^4m_s^4}{12M_B^8})e^{-\frac{m_s^2}{{M_B}^2}}\nonumber\\
&&-8m_sa_s\int_{m_s^2}^{s_0}e^{-\frac{s}{M_B^2}}ds.
\end{eqnarray}
where the notation $x=m_s^2/s$ is used, and the other parameters employed are the standard values: $a=-(2\pi)^2\langle\bar uu\rangle=0.55\; \mbox{GeV}^{3}$,
$b=(2\pi)^2\langle\alpha_sG^2/\pi\rangle=0.47\; \mbox{GeV}^{4}$, $a_s=-(2\pi)^2\langle\bar ss\rangle=0.8a$, and $\langle\bar ug_c\sigma\cdot
Gu\rangle=0.8\langle\bar uu\rangle$.

\begin{figure}
\begin{minipage}{7cm}
\epsfxsize=6cm \centerline{\epsffile{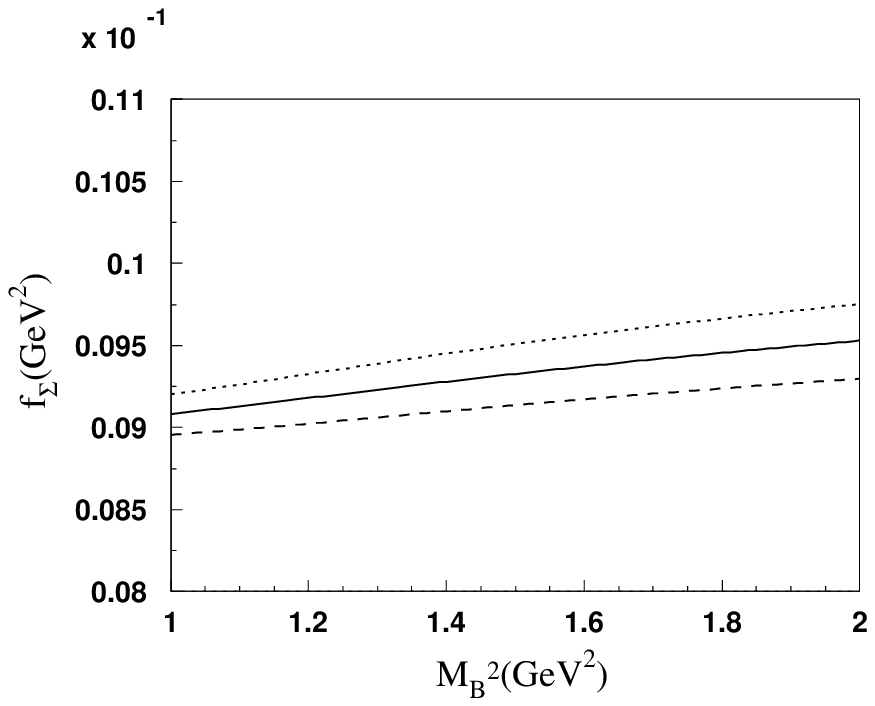}}
\end{minipage}
\hfill
\begin{minipage}{7cm}
\epsfxsize=6cm \centerline{\epsffile{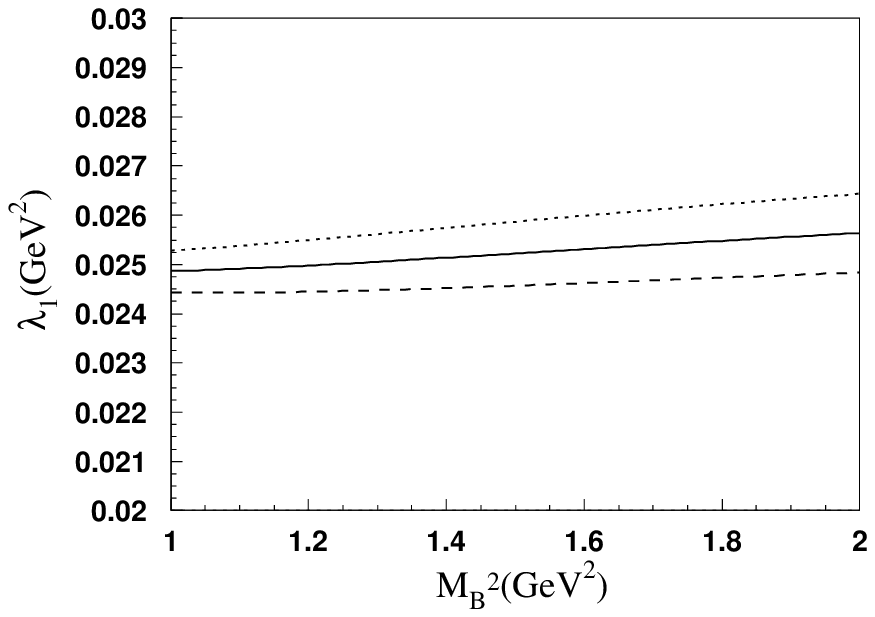}}
\end{minipage}
\hfill
\begin{minipage}{7cm}
\epsfxsize=6cm \centerline{\epsffile{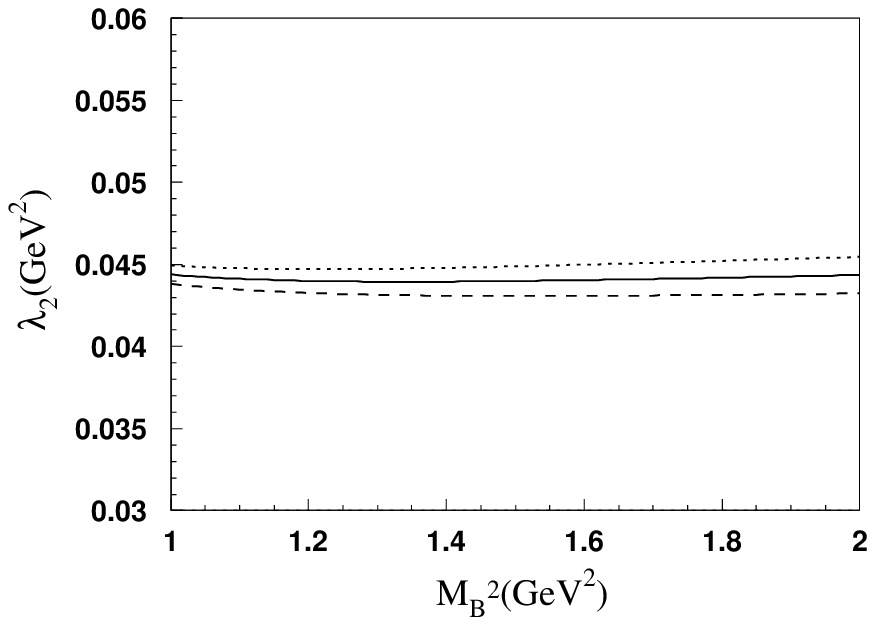}}
\end{minipage}
\hfill
\begin{minipage}{7cm}
\epsfxsize=6cm \centerline{\epsffile{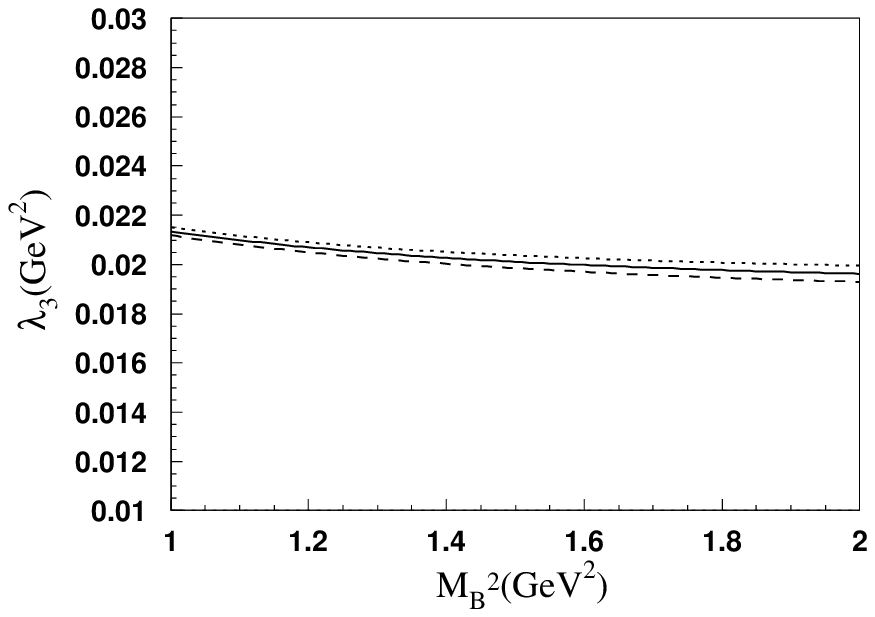}}
\end{minipage}
\caption{}\label{fig1}
\end{figure}

As the usual way of the sum rules, the auxiliary Borel parameter $M_B^2$ should have a proper range in which the results of the sum rules vary mildly with it. On
the one hand the Borel parameter is expected to be large so that the higher order dimensional contributions are suppressed, and on the other hand the Borel
parameter needs to be small enough to suppress the higher resonance contributions. Fig. \ref{fig1} shows the dependence of the above parameters on the Borel
parameter $M_B^2$. The window of Borel parameter is choose as $1\; \mbox{GeV}^2\leq M_B^2\leq 2\; \mbox{GeV}^2$, in which our results are acceptable.

To determine the relative sign of $f_{\Sigma^+}$ and $\lambda_1$, we give the sum rule of $f_{\Sigma^+}\lambda_1^*$:
\begin{eqnarray}
4(2\pi)^4f_{\Sigma^+}\lambda_1^*Me^{-\frac{M^2}{M_B^2}}=-\frac{m_s}{3}\int_{m_s^2}^{s_0}s[(1-x)(3+13x-5x^2+x^3)+12x\ln
x]e^{-\frac{s}{M_B^2}}ds\nonumber\\
-\frac {b}{6}m_s\int_{m_s^2}^{s_0}\frac1s(1-x)[1+\frac{(1-x)(2-5x)}{3x}]e^{-\frac{s}{M_B^2}}ds-\frac{4}{3}a_s\int_{m_s^2}^{s_0}e^{-\frac{s}{M_B^2}}ds.
\end{eqnarray}

\begin{figure}
\begin{minipage}{7cm}
\epsfxsize=6cm \centerline{\epsffile{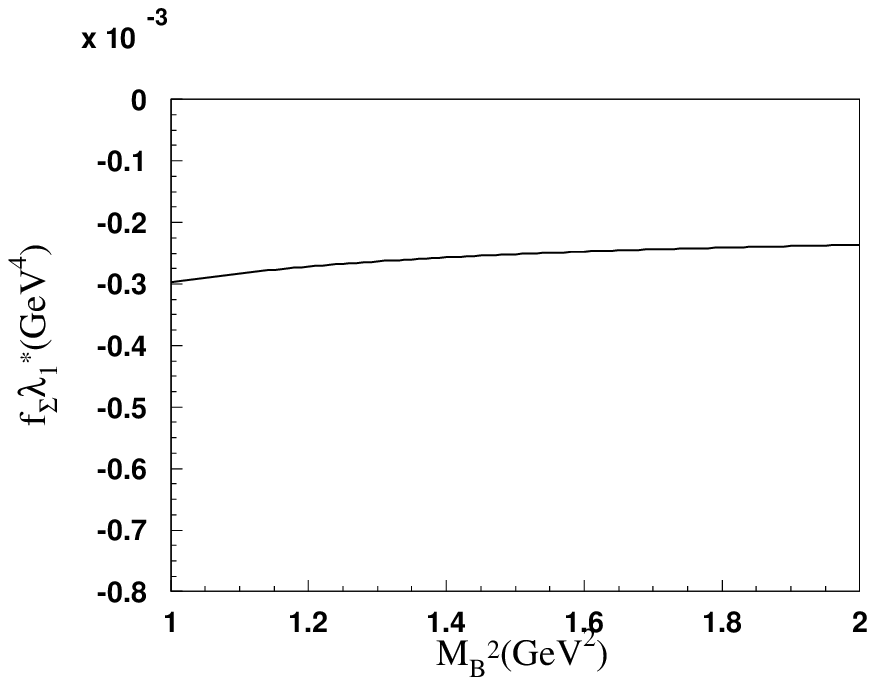}}
\end{minipage}
\hfill
\begin{minipage}{7cm}
\epsfxsize=6cm \centerline{\epsffile{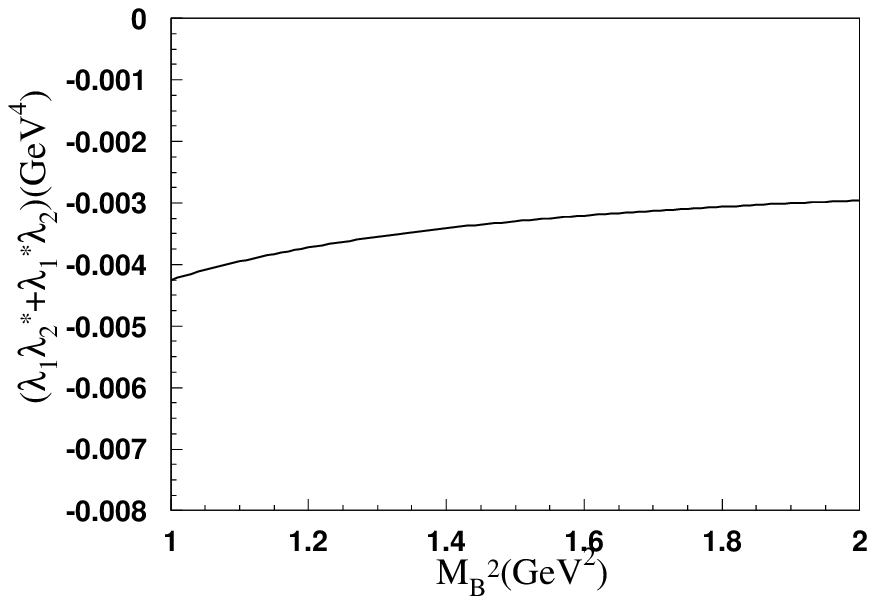}}
\end{minipage}
\hfill
\begin{minipage}{7cm}
\epsfxsize=6cm \centerline{\epsffile{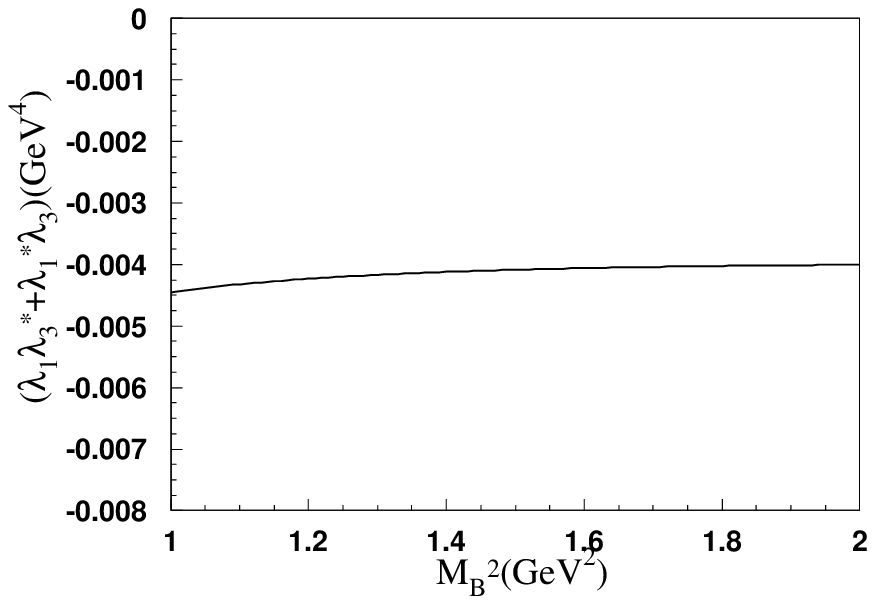}}
\end{minipage}
\caption{}\label{fig2}
\end{figure}

It is similar that the relative sign of $\lambda_2$ and $\lambda_3$ to $\lambda_1$ can be given by the following two sum rules:
\begin{eqnarray}
(2\pi)^4(\lambda_1\lambda_2^*+\lambda_1^*\lambda_2)M^3e^{-\frac{M^2}{{M_B}^2}}=-12m_sa\int_{m_s^2}^{s_0}(1+m_0^2/s)(1-x)^2e^{-\frac{s}{{M_B}^2}}ds,
\end{eqnarray}
and
\begin{eqnarray}
(2\pi)^4(\lambda_1\lambda_3^*+\lambda_1^*\lambda_3)M^3e^{-\frac{M^2}{{M_B}^2}}=-\int_{m_s^2}^{s_0}\{as(1-x)^2(2+x)\nonumber\\
+\frac{m_0^2a}{2}[1-\frac32(1-x)(1+x)+(1-x)(13-25x+2x^2)]\}e^{-\frac{s}{{M_B}^2}}ds.
\end{eqnarray}
 Fig. \ref{fig2} gives the dependence
of the above sum rules on the Borel parameter $M_B^2$.

The final numerical values of the coupling constants of $\Sigma$ are:
\begin{eqnarray}
f_{\Sigma}&=&(9.4\pm0.4)\times10^{-3}\; \mbox{GeV}^2,\hspace{2.5cm}\lambda_1=-(2.5\pm0.1)\times10^{-2}\; \mbox{GeV}^2,\nonumber\\
\lambda_2&=&(4.4\pm0.1)\times10^{-2}\; \mbox{GeV}^2,\hspace{2.5cm}\lambda_3=(2.0\pm0.1)\times10^{-2}\; \mbox{GeV}^2.\label{sigmapara}
\end{eqnarray}

In Ref. \cite{Chernyak}, Chernyak $et$ $al.$ have calculated nonperturbative parameters relevant to leading twist contributions. Their numerical results are
$|f_\Sigma|\simeq 0.51\times10^{-2}\; \mbox{GeV}^2$ and $|f_\Sigma^T|\simeq 0.49\times10^{-2}\; \mbox{GeV}^2$. By contrast, we use Eqs. (\ref{def-nonlocal}) and
(\ref{nonlocalpara}) to give our estimations: $|f_\Sigma|=0.94\times10^{-2}\; \mbox{GeV}^2$ and $|f_\Sigma^T|=0.60\times10^{-2}\; \mbox{GeV}^2$. In comparison with
their results, our predictions are lager in absolute values.
\subsection{QCD sum rules for $\Lambda$ baryon}
\begin{figure}
\begin{minipage}{7cm}
\epsfxsize=6cm \centerline{\epsffile{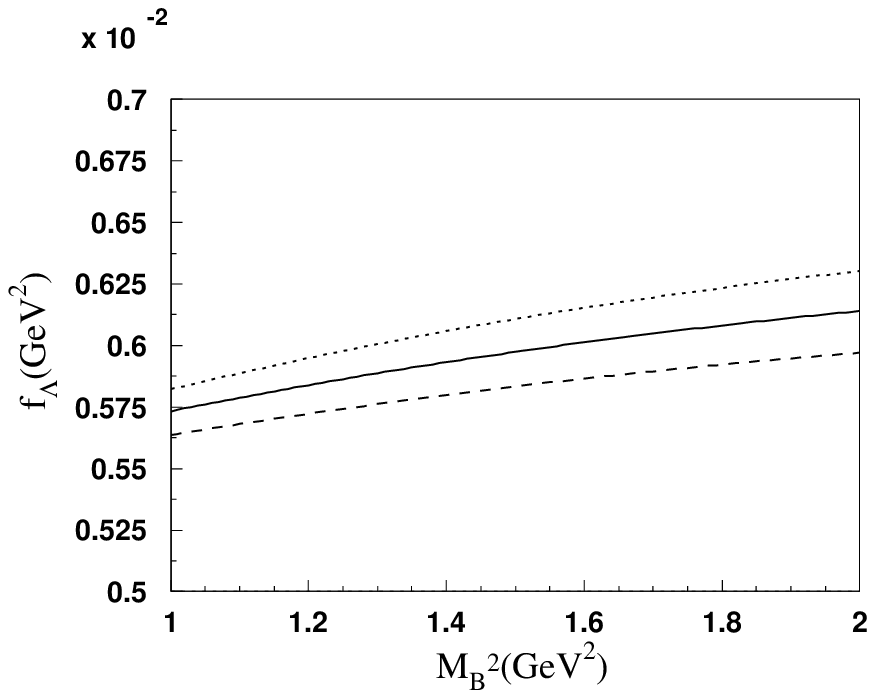}}
\end{minipage}
\hfill
\begin{minipage}{7cm}
\epsfxsize=6cm \centerline{\epsffile{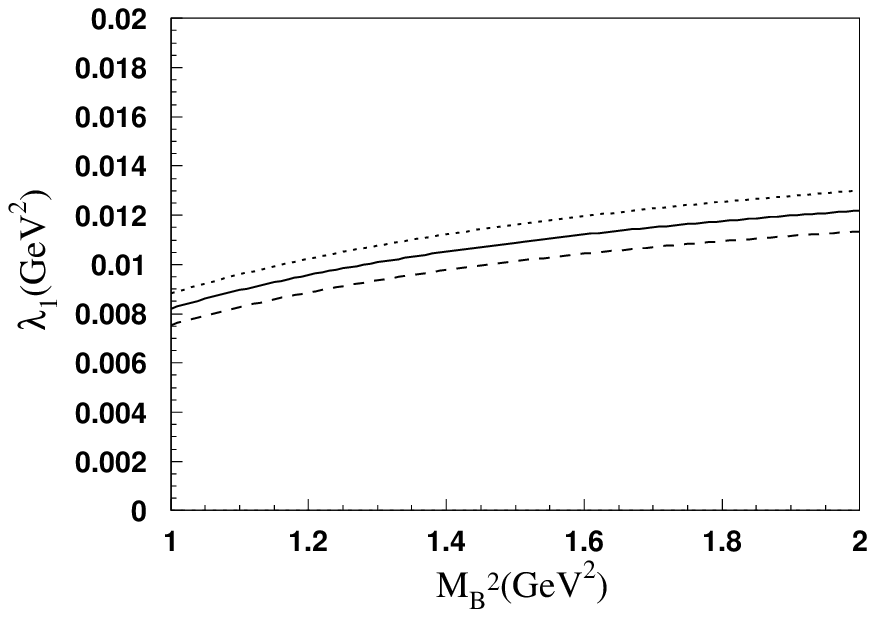}}
\end{minipage}
\hfill
\begin{minipage}{7cm}
\epsfxsize=6cm \centerline{\epsffile{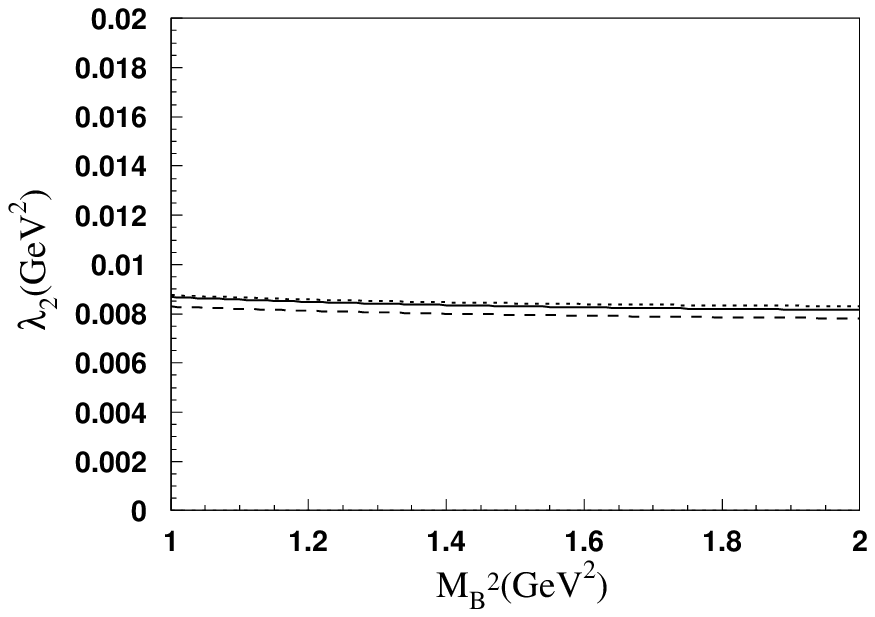}}
\end{minipage}
\hfill
\begin{minipage}{7cm}
\epsfxsize=6cm \centerline{\epsffile{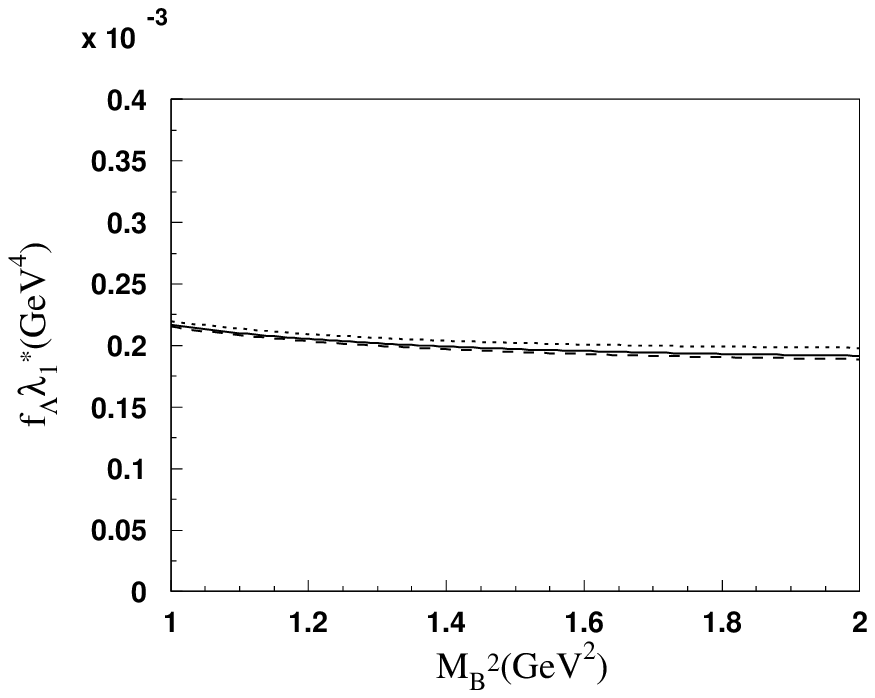}}
\end{minipage}
\caption{}\label{fig3}
\end{figure}

The sum rules of the $\Lambda$ baryon parameters begin with the following correlation functions:
\begin{equation}
\Pi_i(q^2)=\int d^4x e^{iq\cdot x}\langle 0|T\{j_i(x)\bar j_i(0)\}|0\rangle,
\end{equation}
with the definitions of the currents:
\begin{eqnarray}
j_1(x)&=& \epsilon^{ijk} \left[u^i(x) C\gamma_5 \!\not\!{z}d^j(x)\right] \!\not\!{z} s^k(x)\,, \\
j_2(x)&=& \epsilon^{ijk} \left[u^i(x) C\gamma_5\gamma_\mu
d^j(x)\right]\, \gamma^\mu s^k(x) \,,\\
j_3(x)&=& \epsilon^{ijk} \left[u^i(x) C\gamma_5 d^j(x)\right] \, s^k(x) \,,\\
j_4(x)&=& \epsilon^{ijk} \left[u^i(x) Cd^j(x)\right]\,\gamma_5 s^k(x) \,.
\end{eqnarray}
The similar processes as in the above subsection lead to the following results:
\begin{eqnarray}
(4\pi)^4f_\Lambda^2e^{-\frac{M^2}{M_B^2}}=\frac25\int_{m_0^2}^{s_0}s(1-x)^5e^{-\frac{s}{M_B^2}}ds-\frac
b3\int_{m_s^2}^{s_0}\frac1sx(1-x)(1-2x)e^{-\frac{s}{M_B^2}}ds,
\end{eqnarray}
and
\begin{eqnarray}
4(2\pi)^4\lambda_1^2M^2e^{-\frac{M^2}{M_B^2}}&=&\frac12\int_{m_s^2}^{s_0}s^2[(1-x)(1+x)(1-8x+x^2)-12x^2\ln
x]e^{-\frac{s}{M_B^2}}ds\nonumber\\
&&+\frac{b}{12}\int_{m_s^2}^{s_0}(1-x)^2e^{-\frac{s}{M_B^2}}ds-\frac43a^2(1-\frac{m_0^2}{2M_B^2}-\frac{m_0^2m_s^2}{2M_B^4}\nonumber\\
&& +\frac{m_0^4m_s^4}{16M_B^8})e^{-\frac{m_s^2}{M_B^2}}-m_sa_s\int_{m_s^2}^{s_0}e^{-\frac{s}{M_B^2}}ds,
\end{eqnarray}
and
\begin{eqnarray}
4(4\pi)^4\lambda_2^2M^2e^{-\frac{M^2}{M_B^2}}&=&\int_{m_s^2}^{s_0}s^2[(1-x)(1+x)(1-8x+x^2)-12x^2\ln
x]e^{-\frac{s}{M_B^2}}ds\nonumber\\
&&+\frac{b}{3}\int_{m_s^2}^{s_0}(1-x)(1+5x)e^{-\frac{s}{M_B^2}}ds+\frac{32}{3}a^2(1-\frac{m_0^2}{2M_B^2}-\frac{m_0^2m_s^2}{2M_B^4}
\nonumber\\
&&+\frac{m_0^4m_s^4}{16M_B^8})e^{-\frac{m_s^2}{M_B^2}}-4m_sa_s\int_{m_s^2}^{s_0}e^{-\frac{s}{M_B^2}}ds,
\end{eqnarray}
and
\begin{eqnarray}
4(4\pi)^4\lambda_3^2M^2e^{-\frac{M^2}{M_B^2}}&=&\int_{m_s^2}^{s_0}s^2[(1-x)(1+x)(1-8x+x^2)-12x^2\ln
x]e^{-\frac{s}{M_B^2}}ds\nonumber\\
&&+\frac{b}{3}\int_{m_s^2}^{s_0}(1-x)(1+5x)e^{-\frac{s}{M_B^2}}ds+\frac{32}{3}a^2(1-\frac{m_0^2}{2M_B^2}-\frac{m_0^2m_s^2}{2M_B^4}\nonumber\\
&&+\frac{m_0^4m_s^4}{16M_B^8})e^{-\frac{m_s^2}{M_B^2}}-4m_sa_s\int_{m_s^2}^{s_0}e^{-\frac{s}{M_B^2}}ds,
\end{eqnarray}
and the sum rule of $f_\Lambda\lambda_1^*$ is
\begin{eqnarray}
(4\pi)^4f_\Lambda\lambda_1^*Me^{-\frac{M^2}{M_B^2}}=\frac23m_s\int_{m_s^2}^{s_0}s[(1-x)(3+13x-5x^2+x^3)+12x\ln
x]e^{-\frac{s}{M_B^2}}ds\nonumber\\
+\frac b3m_s\int_{m_s^2}^{s_0}\frac1s(1-x)[1+\frac{(1-x)(2-5x)}{3x}]e^{-\frac{s}{M_B^2}}ds+\frac{8}{3}a_s\int_{m_s^2}^{s_0}e^{-\frac{s}{M_B^2}}ds.
\end{eqnarray}
Note that the sum rules of $\lambda_2$ and $\lambda_3$ are the same. In Fig. \ref{fig3}, the sum rules of the parameters on the Borel parameter $M_B^2$ are shown.
The final numerical results for the parameters of $\Lambda$ are:
\begin{eqnarray}
f_\Lambda&=&(6.0\pm0.3)\times10^{-3}\; \mbox{GeV}^2,\hspace{2.5cm}\lambda_1=(1.0\pm0.3)\times10^{-2}\; \mbox{GeV}^2,\nonumber\\
|\lambda_2|&=&(0.83\pm0.05)\times10^{-2}\; \mbox{GeV}^2,\hspace{2.0cm}|\lambda_3|=(0.83\pm0.05)\times10^{-2}\; \mbox{GeV}^2.\label{lambdapara}
\end{eqnarray}
In the above results, $f_\Lambda$ and $\lambda_1$ have the same sign, which is different from that shown in Ref. \cite{Wang}. The relative sign of $\lambda_1$ and
$\lambda_2$ cannot be determined by the method presented above. Here we only list the absolute values of the two parameters.

The numerical estimations by Chernyak $et$ $al.$ are: $|f_\Lambda|\simeq 0.63\times10^{-2}\; \mbox{GeV}^2$ and $|f_\Lambda^T|\simeq 0.063\times10^{-2}\;
\mbox{GeV}^2$. Our result on $f_\Lambda$ is $f_\Lambda=0.60\times10^{-2}\; \mbox{GeV}^2$, but in our calculations $f_\Lambda^T$ is zero, which is different from
that of Chernyak.  The deviation is due to the tensor structure of the baryon, which disappears at the leading order of the conformal expansion because of the
isospin symmetry. In our approach, the determination of the tensor coupling constant relies on the next conformal expansion of the DAs, which corresponds to the sum
rules of higher order moments in Ref. \cite{Chernyak}.
\section{Application:electromagnetic form factors of the baryons with light-cone QCD
sum rules}\label{sec:app}
\subsection{LCSR for the electromagnetic form factors}
The EM form factors of hadrons are fundamental objects for understanding their internal structures. There were a lot of investigations on various hadrons both
experimentally and theoretically, including meson \cite{Bebek,Dally,Liesenfeld,Volmer,Horn,Tadevosyan} and baryon
\cite{Walker,Andivahis,Arrington,Christy,Bosted,Lung,Qattan,Bourgeois,Anklin,Kubon}. While as there were few experimental data and theoretical investigations, the
EM form factors of the baryons, such as $\Sigma$ and $\Lambda$ and so on, have not received much attention in the past years. Chiral perturbation theory and the
chiral quark/soliton model have been used to study the baryon EM form factors at low momentum transfer \cite{Kim,Kubis}. T. Van Cauteren $et$ $al.$ have
investigated the electric and magnetic form factors of these baryons in the relativistic constituent quark model \cite{Cauteren}. In the previous work
\cite{LambdaEM} we gave an investigation on the EM form factors of the $\Lambda$ baryon. In this section, the EM form factors of the $\Sigma$ baryon are
investigated at moderately large momentum transfer within the framework of light-cone QCD sum rule method, and the magnetic moments of the same baryons are
estimated by comparing our results with the existing dipole formula.

The matrix element of the electromagnetic current between the baryon states can be expressed as the Dirac and Pauli form factors $F_1(Q^2)$ and $F_2(Q^2)$,
respectively:
\begin{equation}
\langle\Sigma(P,s)|j_\mu^{em}(0)|\Sigma(P',s')\rangle=\bar \Sigma (P,s)[\gamma_\mu F_1(Q^2)-i\frac{\sigma_{\mu\nu} q^\nu}{2M}F_2(Q^2)]\Sigma(P',s'),\label{form}
\end{equation}
where $j_\mu^{em}=e_u\bar u\gamma_\mu u+e_s \bar s\gamma_\mu s$ is the electromagnetic current relevant to the hadron, and $P,s$ and $P',s'$ are the four-momenta
and the spins of the initial and the final $\Sigma$ baryon states, respectively. From the experimental viewpoint, the EM form factors can be expressed by the
electric $G_E(Q^2)$ and magnetic $G_M(Q^2)$ Sachs form factors:
\begin{eqnarray}
G_M(Q^2)&=&F_1(Q^2)+F_2(Q^2),\nonumber\\
G_E(Q^2)&=&F_1(Q^2)-\frac{Q^2}{4M^2}F_2(Q^2),
\end{eqnarray}
and at the point $Q^2=0$ the magnetic $G_M(Q^2)$ form factor gives the magnetic moment of the baryon:
\begin{equation}
G_M(0)=\mu_\Sigma.
\end{equation}
In order to evaluate the magnetic moment of the baryon from its EM form factors, the magnetic form factor $G_M(Q^2)$ is assumed to be described by the dipole
formula:
\begin{equation}
\frac{1}{\mu_\Sigma}G_M(Q^2)=\frac{1}{(1+Q^2/m_0^2)^2}=G_D(Q^2).\label{dipole}
\end{equation}
As there is no information about the parameter $m_0^2$ from experimental data, the two parameters $m_0^2$ and $\mu_\Sigma$ are estimated simultaneously by fitting
the magnetic form factor with the dipole formula (\ref{dipole}).

The calculation mainly focuses on $\Sigma^+$ baryon and it is similar for the calculation of $\Sigma^-$. The process of the derivation begins with the correlation
function:
\begin{equation}
T_\mu(P,q)=i \int d^4xe^{iq\cdot x}\langle 0|T\{j_{\Sigma^+}(0)j_\mu^{em}(x)\}|\Sigma^+(P,s)\rangle,\label{correlator}
\end{equation}
where the interpolating current $j_{\Sigma^+}$ is chosen as Eq. (\ref{CZcurrent}). The hadronic representation of the correlation function is acquired by inserting
a complete set of states with the same quantum numbers as those of $\Sigma^+$:
\begin{eqnarray}
z^\mu T_\mu(P,q)&=&\frac{1}{M^2-P'^2}f_{\Sigma^+} (P'\cdot z)[2(P'\cdot zF_1(Q^2)-\frac{q\cdot z}{2}F_2(Q^2))\!\not\! z\nonumber\\&&+(P'\cdot zF_2(Q^2)+\frac{q\cdot
z}{2}F_2(Q^2))\frac{\!\not\! z\!\not\! q}{M}]\Sigma^+(P,s)+...,
\end{eqnarray}
where $P'=P-q$, and the dots stand for the higher resonances and continuum contributions. The correlation function is contracted with $z^\mu$ to get rid of
contributions proportional to $z^\mu$ which is subdominant on the light cone. On the theoretical side, the correlation function (\ref{correlator}) can be calculated
to leading order of $\alpha_s$ as
\begin{eqnarray}
z_\mu T^\mu &=&(P\cdot z)^2(\!\not\! {z}\Sigma^+)_\gamma \Big\{4e_u\int_0^1d\alpha_2\frac{1}{s-p'^2}\{B_0(\alpha_2)+\frac{M^2}{(s-p'^2)}B_1(\alpha_2)
\nonumber\\
&&-2\frac{M^4}{(s-P'^2)^2}B_2(\alpha_2)\}+2e_s\int_0^1d\alpha_3\frac{1}{s_2-P'^2}\{C_0(\alpha_3)
\nonumber\\
&&+\frac{M^2}{(s_2-P'^2)}C_1(\alpha_3)-2\frac{M^4}{(s_2-P'^2)^2}C_2(\alpha_3)\} \Big\}\nonumber\\
&&+(P\cdot
z)^2M(\!\not\!{z}\!\not\!{q}\Sigma^+)_\gamma\Big\{4e_u\int_0^1d\alpha_2\frac{1}{\alpha_2(s-P'^2)^2}\{-D_1(\alpha_2)\nonumber\\
&&+2\frac{M^2}{(s-P'^2)}B_2(\alpha_2)\}+e_s\int_0^1d\alpha_3\frac{1}{\alpha_3(s_2-P'^2)^2}\{-E_1(\alpha_3)\nonumber\\
&& +2\frac{M^2}{(s_2-P'^2)}C_2(\alpha_3)\}\Big\},
\end{eqnarray}
where $s=(1-\alpha_2)M^2+\frac{(1-\alpha_2)}{\alpha_2}Q^2$ and $s_2=(1-\alpha_3)M^2+\frac{(1-\alpha_3)}{\alpha_3}Q^2+\frac{m_s^2}{\alpha_3}$. Here the following
notations are used for convenience:
\begin{eqnarray}
B_0(\alpha_2)&=&\int_0^{1-\alpha_2}d\alpha_1V_1(\alpha_1,\alpha_2,1-\alpha_1-\alpha_2),\nonumber\\
B_1(\alpha_2)&=&(2\widetilde V_1-\widetilde V_2-\widetilde V_3-\widetilde V_4-\widetilde V_5)(\alpha_2),\nonumber\\
B_2(\alpha_2)&=&(-\widetilde{\widetilde V_1}+\widetilde{\widetilde V_2}+\widetilde{\widetilde V_3}+\widetilde{\widetilde V_4}+\widetilde{\widetilde
V_5}-\widetilde{\widetilde
V_6})(\alpha_2),\nonumber\\
C_0(\alpha_3)&=&\int_0^{1-\alpha_3}d\alpha_1V_1(\alpha_1,1-\alpha_1-\alpha_3,\alpha_3),\nonumber\\
C_1(\alpha_3)&=&(2\widetilde V_1-\widetilde V_2-\widetilde V_3-\widetilde V_4-\widetilde V_5)(\alpha_3),\nonumber\\
C_2(\alpha_3)&=&(-\widetilde{\widetilde V_1}+\widetilde{\widetilde V_2}+\widetilde{\widetilde V_3}+\widetilde{\widetilde V_4}+\widetilde{\widetilde
V_5}-\widetilde{\widetilde
V_6})(\alpha_3),\nonumber\\
D_1(\alpha_2)&=&(\widetilde V_1-\widetilde V_2-\widetilde V_3)(\alpha_2),\nonumber\\
E_1(\alpha_3)&=&(\widetilde V_1-\widetilde V_2-\widetilde V_3)(\alpha_3),
\end{eqnarray}
where
\begin{eqnarray}
\widetilde V_i(\alpha_2)&=&\int_0^{\alpha_2}d{\alpha_2'}\int_0^{1-\alpha_2'}d\alpha_1V_i(\alpha_1,\alpha_2',1-\alpha_1-\alpha_2'),\nonumber\\
\widetilde{\widetilde V_i}(\alpha_2)&=&\int_0^{\alpha_2}d{\alpha_2'}\int_0^{\alpha_2'}d{\alpha_2''}\int_0^{1-\alpha_2''}d\alpha_1V_i(\alpha_1,\alpha_2'',1-\alpha_1-\alpha_2''),\nonumber\\
\widetilde V_i(\alpha_3)&=&\int_0^{\alpha_3}d{\alpha_3'}\int_0^{1-\alpha_3'}d\alpha_1V_i(\alpha_1,1-\alpha_1-\alpha_3',\alpha_3'),\nonumber\\
\widetilde{\widetilde
V_i}(\alpha_3)&=&\int_0^{\alpha_3}d{\alpha_3'}\int_0^{\alpha_3'}d{\alpha_3''}\int_0^{1-\alpha_3''}d\alpha_1V_i(\alpha_1,1-\alpha_1-\alpha_3'',\alpha_3'').
\end{eqnarray}

Then equating both sides of the Borel transformed version of hadronic and theoretical representations with the assumption of quark-hadron duality, the final sum
rules are given as follows:
\begin{eqnarray}
2f_{\Sigma^+}F_1(Q^2)e^{-\frac{M^2}{{M_B}^2}}&=&4e_u\int_{\alpha_{20}}^1d\alpha_2e^{-\frac{s}{{M_B}^2}}\Big\{B_0(\alpha_2)
+\frac{M^2}{{M_B}^2}B_1(\alpha_2)-\frac{M^4}{{M_B}^4}B_2(\alpha_2)\Big\}\nonumber\\
&&+4e_ue^{-\frac{s_0}{{M_B}^2}}\frac{\alpha_{20}^2M^2}{\alpha_{20}^2M^2+Q^2}\Big\{B_1(\alpha_{20})-\frac{M^2}{M_B^2}B_2(\alpha_{20})\Big\}\nonumber\\
&&+4e_ue^{-\frac{s_0}{{M_B}^2}}\frac{\alpha_{20}^2M^4}{\alpha_{20}^2M^2+Q^2}\frac{d}{d\alpha_{20}}
B_2(\alpha_{20})\frac{\alpha_{20}^2}{\alpha_{20}^2M^2+Q^2}\nonumber\\
&&+2e_s\int_{\alpha_{30}}^1d\alpha_3e^{-\frac{s_2}{{M_B}^2}}\Big\{C_0(\alpha_3)
+\frac{M^2}{{M_B}^2}C_1(\alpha_3)-\frac{M^4}{{M_B}^4}C_2(\alpha_3)\Big\}\nonumber\\
&&+2e_se^{-\frac{s_0}{{M_B}^2}}\frac{\alpha_{30}^2M^2}{\alpha_{30}^2M^2+Q^2+m_s^2}\Big\{C_1(\alpha_{30})-\frac{M^2}{M_B^2}C_2(\alpha_{30})\Big\}\nonumber\\
&&+2e_se^{-\frac{s_0}{{M_B}^2}}\frac{\alpha_{30}^2M^4}{\alpha_{30}^2M^2+Q^2+m_s^2}\frac{d}{d\alpha_{30}}
C_2(\alpha_{30})\frac{\alpha_{30}^2}{\alpha_{30}^2M^2+Q^2+m_s^2}\nonumber\\
\end{eqnarray}
for $F_1(Q^2)$ and
\begin{eqnarray}
f_{\Sigma^+}F_2(Q^2)e^{-\frac{M^2}{{M_B}^2}}&=&M^2\Big\{4e_u\int_{\alpha_{20}}^1d\alpha_2e^{-\frac{s}{{M_B^2}}}\frac{1}{\alpha_2{M_B}^2}\big\{-D_1(\alpha_2)+\frac{M^2}{{M_B}^2}B_2(\alpha_2)\big\}
\nonumber\\&&-4e_ue^{-\frac{s_0}{{M_B}^2}}\frac{\alpha_{20}}{\alpha_{20}^2M^2+Q^2}\big\{D_1(\alpha_{20})-\frac{M^2}{M_B^2}B_2(\alpha_{20})\big\}\nonumber\\
&&-4e_ue^{-\frac{s_0}{{M_B}^2}}\frac{\alpha_{20}^2M^2}{\alpha_{20}^2M^2+Q^2}
\frac{d}{d\alpha_{20}}B_2(\alpha_{20})\frac{\alpha_{20}}{\alpha_{20}^2M^2+Q^2}\nonumber\\
&&+2e_s\int_{\alpha_{30}}^1d\alpha_{30}e^{-\frac{s_2}{{M_B}^2}}\frac{1}{\alpha_3{M_B}^2}\big\{-E_1(\alpha_3)+\frac{M^2}{{M_B}^2}C_2(\alpha_3)\big\}\nonumber\\
&&-2e_se^{-\frac{s_{0}}{M_B^2}}\frac{\alpha_{30}}{\alpha_{30}^2M^2+Q^2+m_s^2}\big\{E_1(\alpha_{30})-\frac{M^2}{M_B^2}C_2(\alpha_{30})\big\}\nonumber\\
&&-2e_se^{-\frac{s_{0}}{M_B^2}}\frac{\alpha_{30}^2M^2}{\alpha_{30}^2M^2+Q^2+m_s^2}\frac{d}{d\alpha_{30}}C_2(\alpha_{30})
\frac{\alpha_{30}}{\alpha_{30}^2M^2+Q^2+m_s^2}\Big \}\nonumber\\
\end{eqnarray}
for $F_2(Q^2)$.
\subsection{Numerical analysis}
\begin{figure}
\begin{minipage}{7cm}
\epsfxsize=7cm \centerline{\epsffile{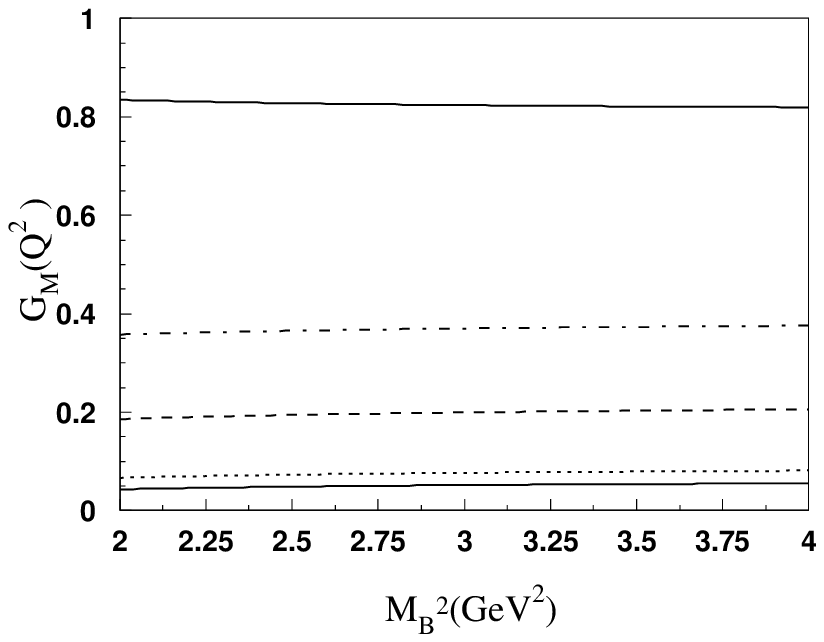}}
\end{minipage}
\hfill \begin{minipage}{7cm} \epsfxsize=7cm
\centerline{\epsffile{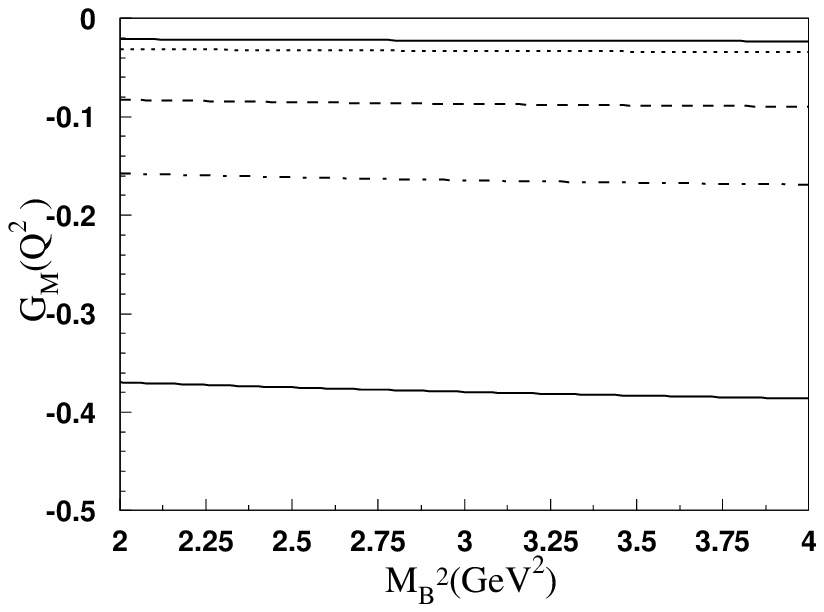}}
\end{minipage}
\caption{}\label{fig4}
\end{figure}

In the numerical analysis, the continuum threshold is chosen as $s_0=(2.65-2.85)\, \mbox{GeV}^2$. The masses of the $\Sigma$ baryons from Ref. \cite{PDG} are
$M_{\Sigma^+}=1.189\;\ \mbox{GeV}$ and $M_{\Sigma^-}=1.197\;\ \mbox{GeV}$. The parameters $f_{\Sigma}$ and $\lambda_1$ are used as the central values in Eqs.
(\ref{sigmapara}). For the auxiliary Borel parameter $M_B^2$, there should be a region where the sum rules are almost independent of it. To choose a platform for
the Borel parameter, we should suppress both resonance contributions and the higher twist contributions simultaneously. Fig. \ref{fig4} shows the dependence of the
magnetic form factors on the Borel parameter at different points of $Q^2$. Our results are acceptable in the range $2.0\;\ \mbox{GeV}^2\leq M_B^2\leq 4.0\;\
\mbox{GeV}^2$.

\begin{figure}
\begin{minipage}{7cm}
\epsfxsize=6cm \centerline{\epsffile{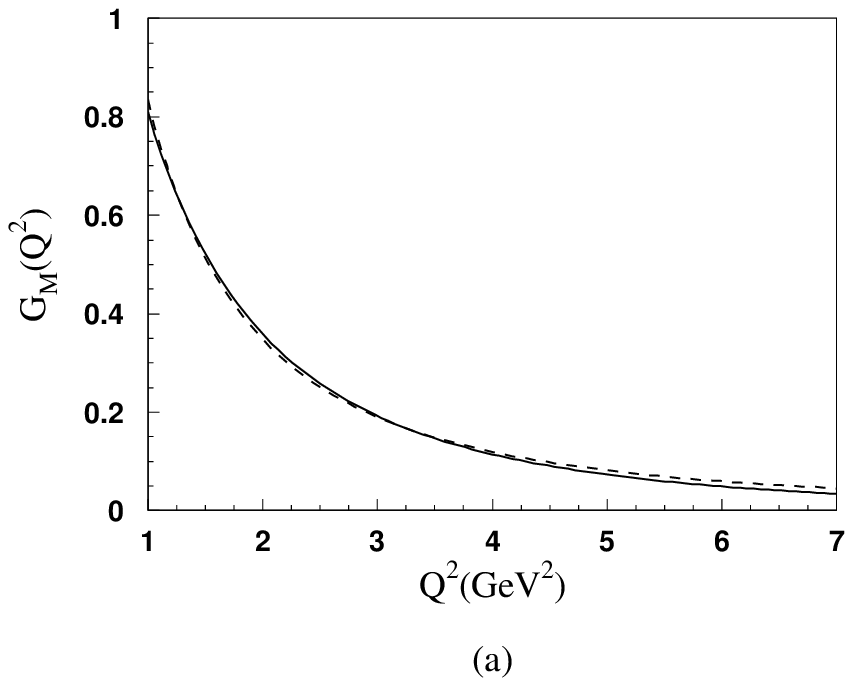}}
\end{minipage}
\hfill
\begin{minipage}{7cm}
\epsfxsize=6cm \centerline{\epsffile{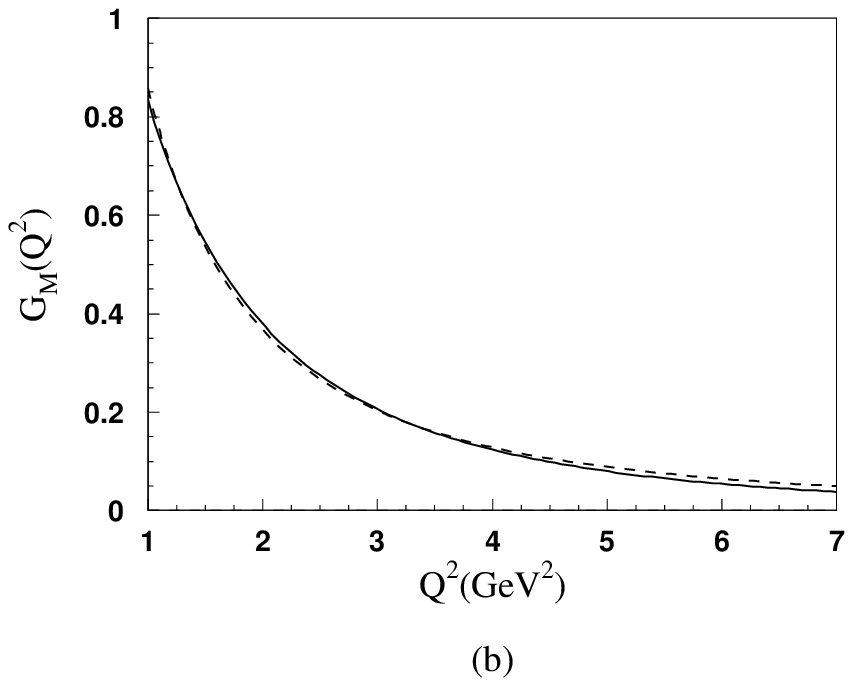}}
\end{minipage}
\caption{}\label{fig5}
\end{figure}

The estimation on the magnetic moment of the baryon comes from the fitting of the magnetic form factor by the dipole formula (\ref{dipole}). In the following
analysis the Borel parameter is chosen to be $M_B^2=3\;\ \mbox{GeV}^2$. Fig. \ref{fig5} gives the dependence of the magnetic form factor $G_M(Q^2)$ on the momentum
transfer at different points of the threshold $s_0$. The figure shows that $G_M(Q^2)$ decreases with $Q^2$, which is in accordance with the assumption in Eq.
(\ref{dipole}). To estimate the magnetic moment numerically, the magnetic form factor $G_M(Q^2)$ is fitted by the formula $\mu_\Sigma/(1+Q^2/m_0^2)^2$, which is
described by the dashed lines in Fig. \ref{fig5}. From the figures the magnetic moment of $\Sigma^+$ is given as $\mu_{\Sigma^+}=(3.13\pm 0.10)\mu_N$, and the
estimation of the other parameter is $m_0^2=(0.86\pm 0.04) \; \mbox{GeV}^2$.

\begin{figure}
\begin{minipage}{7cm}
\epsfxsize=6cm \centerline{\epsffile{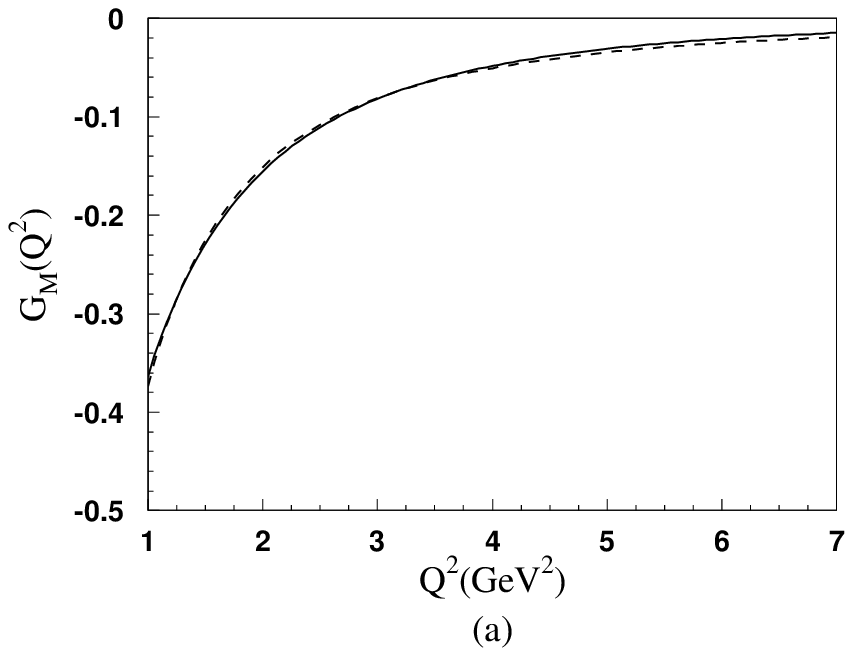}}
\end{minipage}
\hfill
\begin{minipage}{7cm}
\epsfxsize=6cm \centerline{\epsffile{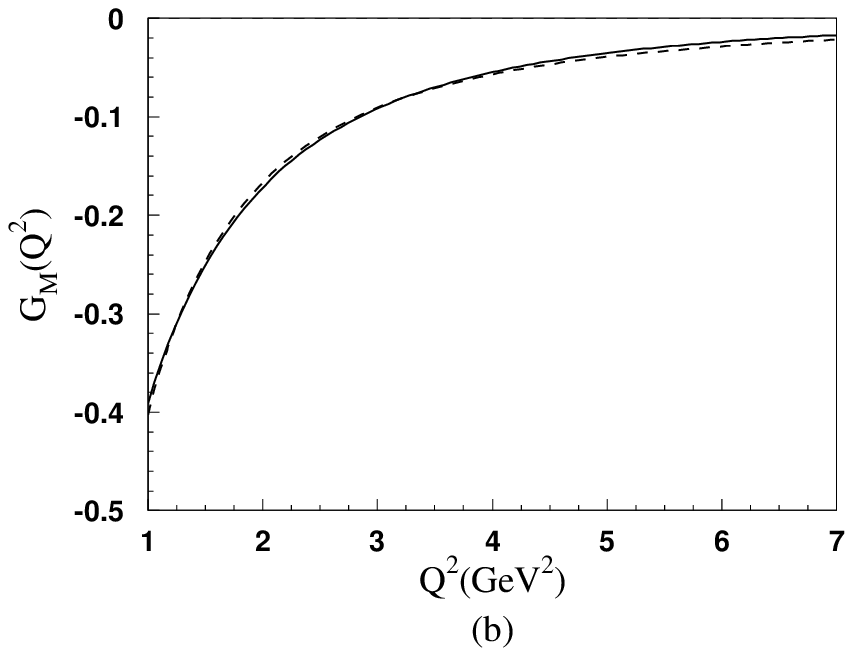}}
\end{minipage}
\caption{}\label{fig6}
\end{figure}

The similar process is carried out for the numerical analysis of $\Sigma^-$. The estimation of the $\Sigma^-$ magnetic moment is shown in Fig. \ref{fig6}, which are
the magnetic form factors of the baryon at different threshold and the fittings by the dipole formula. The numerical values from the fittings are
$\mu_{\Sigma^-}=-(1.59\pm 0.02)\mu_N$ and $m_0^2=(0.78\pm 0.03)\; \mbox{GeV}^2$.

Table \ref{moment} lists magnetic moments of the two baryons from various approaches: data from Particle Data Group (PDG) \cite{PDG}; QCD sum rules \cite{Chiu}
(SR(1) for $\chi=-3.3$ and SR(2) for $\chi=-4.5$); QCD string approach (QCDSA) \cite{Kerbikov}; chiral perturbation theory ($\chi$PT) \cite{Puglia}; Skyrme model
(SKRM) \cite{Park}; light cone sum rules \cite{Aliev2} (LCSR(1) for $\chi=-3.3$ and LCSR(2) for $\chi=-4.5$). The table shows that our results are larger in
absolute values than the others. This may partly lie in the fact that more detailed information on DAs calls for higher order conformal expansion. At the same time,
the choice of the interpolating currents may affect the results to some extent \cite{ff,Aliev}. The estimation is expected to be better if more information about
the DAs are known and higher order QCD coupling $O(\alpha_s)$ effect are included.

\begin{figure}
\begin{minipage}{7cm}
\epsfxsize=6cm \centerline{\epsffile{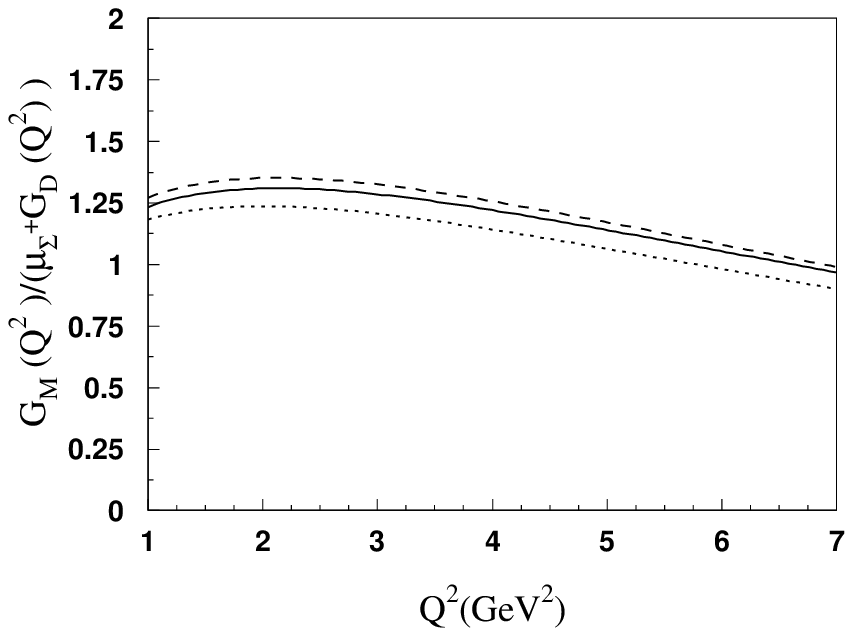}}
\end{minipage}
\hfill
\begin{minipage}{7cm}
\epsfxsize=6cm \centerline{\epsffile{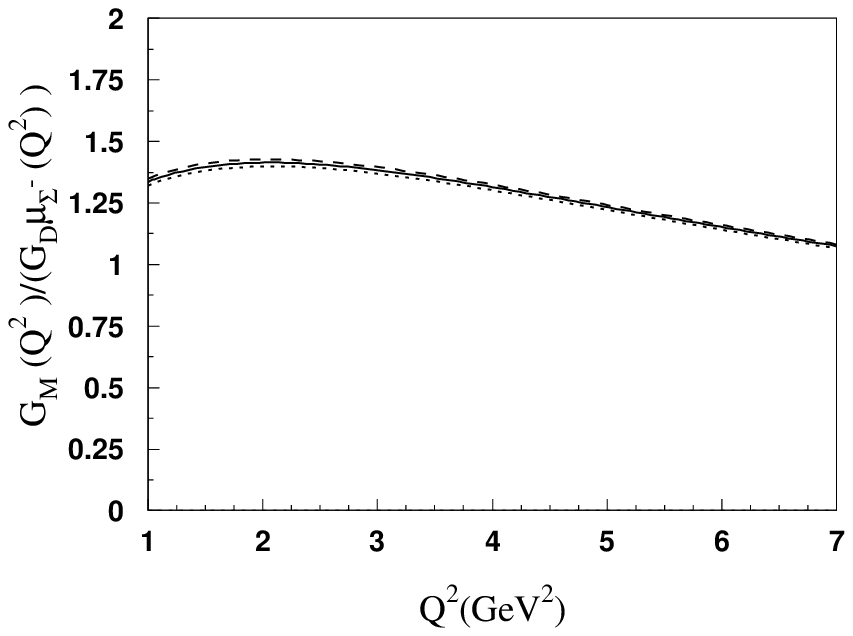}}
\end{minipage}
\caption{}\label{fig7}
\end{figure}

Finally, the $Q^2$-dependence of the physical value $ G_M/{\mu_\Sigma G_D}$ is given in Fig. \ref{fig7}. In the numerical analysis, the input parameters $m_0^2$
used in the dipole formula (\ref{dipole}) are the central values obtained above, which are $m_0^2=0.86\; \mbox{GeV}^2$ for $\Sigma^+$ and $m_0^2=0.78\;
\mbox{GeV}^2$ for $\Sigma^-$, while the magnetic moments used come from Ref. \cite{PDG}, which are $\mu_{\Sigma^+}=2.458\mu_N$ and $\mu_{\Sigma^-}=-1.160\mu_N$.

\section{Summary}\label{sec:sum}
In this paper we present the DAs of baryons with quantum number $I(J^P)=1(\frac{1}{2}^+)$ (for $\Sigma^{\pm}$) and $I(J^P)=0(\frac{1}{2}^+)$ (for $\Lambda$) up to
twist $6$. We find that fourteen independent DAs are needed to describe the structure of the baryons. The method employed is based on the conformal partial wave
expansion, and the nonlocal nonperturbative parameters are determined in the QCD sum rule framework. Our calculation on the conformal expansion of the DAs is to
leading order conformal spin accuracy. Compared with the previous work \cite{Wang}, the calculation on the $\Lambda$ baryon gives DAs of all other Lorentz
structures besides axial-like vector structures. Another new result is that the relative sign of the two parameters $f_\Lambda$ and $\lambda_1$ are positive.

With the DAs obtained, the EM form factors of $\Sigma$ are investigated in the range $1\; \mbox{GeV}^2\leq Q^2\leq 7\; \mbox{GeV}^2$. We assume that the magnetic
form factor can be described by the dipole formula. Fitting the result by the dipole formula, the magnetic moments of the baryons are estimated, which are
$\mu_{\Sigma^+}=(3.13\pm 0.10)\mu_N$, and $\mu_{\Sigma^-}=-(1.59\pm 0.02)\mu_N$. Compared with values given by Particle Data Group \cite{PDG}, our results are
larger in absolute values. This shows that the calculation needs more detailed information on the DAs, which may come from higher order conformal spin
contributions, and at the same time the choice of the interpolating currents may also affect our estimation to some extent.

\section*{Acknowledgments}  This work was supported in part by the National
Natural Science Foundation of China under Contract No.10675167.
\appendix
\section*{Appendices}
In the appendices we give our results on the DAs of $\Sigma$ and $\Lambda$ explicitly. As the definition in Eq. (\ref{da-deftwist}), our results are listed in the
following subsections.
\section{DAs of the $\Sigma$ baryon}
Twist-$3$ distribution amplitudes of $\Sigma$ are:
\begin{eqnarray}
V_1(x_i)&=&120x_1x_2x_3\phi_3^0\,,\hspace{2.5cm}A_1(x_i)=0\,,\nonumber\\
T_1(x_i)&=&120x_1x_2x_3\phi_3^{'0}\,.
\end{eqnarray}
Twist-$4$ distribution amplitudes are:
\begin{eqnarray}
S_1(x_i)&=&6(x_2-x_1)x_3(\xi_4^0+\xi_4^{'0})\,,\hspace{1.6cm}P_1(x_i)=6(x_2-x_1)x_3(\xi_4^0-\xi_4^{'0})\,,\nonumber\\
V_2(x_i)&=&24x_1x_2\phi_4^0\,,\hspace{3.9cm}A_2(x_i)=0\,,\nonumber\\
V_3(x_i)&=&12x_3(1-x_3)\psi_4^0\,,\hspace{2.8cm}A_3(x_i)=-12x_3(x_1-x_2)\psi_4^0\,,\nonumber\\
T_2(x_i)&=&24x_1x_2\phi_4^{'0}\,,\hspace{3.9cm}T_3(x_i)=6x_3(1-x_3)(\xi_4^0+\xi_4^{'0})\,,\nonumber\\
T_7(x_i)&=&6x_3(1-x_3)(\xi_4^{'0}-\xi_4^0)\,.
\end{eqnarray}
Twist-$5$ distribution amplitudes are:
\begin{eqnarray}
S_2(x_i)&=&\frac32(x_1-x_2)(\xi_5^0+\xi_5^{'0})\,,\hspace{1.5cm}P_2(x_i)=\frac32(x_1-x_2)(\xi_5^0-\xi_5^{'0})\,,\nonumber\\
V_4(x_i)&=&3(1-x_3)\psi_5^0\,,\hspace{2.95cm}A_4(x_i)=3(x_1-x_2)\psi_5^0\,,\nonumber\\
V_5(x_i)&=&6x_3\phi_5^0\,,\hspace{4.05cm}A_5(x_i)=0\,,\nonumber\\
T_4(x_i)&=&-\frac32(x_1+x_2)(\xi_5^{'0}+\xi_5^0)\,,\hspace{1.25cm}T_5(x_i)=6x_3\phi_5^{'0}\,,\nonumber\\
T_8(x_i)&=&\frac32(x_1+x_2)(\xi_5^{'0}-\xi_5^0)\,.
\end{eqnarray}
Finally twist-$6$ distribution amplitudes are:
\begin{eqnarray}
V_6(x_i)&=&2\phi_6^0\,,\hspace{2.5cm}A_6(x_i)=0\,,\nonumber\\
T_6(x_i)&=&2\phi_6^{'0}\,.
\end{eqnarray}
\section{DAs of the $\Lambda$ baryon}
Twist-$3$ distribution amplitudes of $\Lambda$ are:
\begin{eqnarray}
V_1(x_i)&=&0\,,\hspace{4.5cm}A_1(x_i)=-120x_1x_2x_3\phi_3^0\,,\nonumber\\
T_1(x_i)&=&0\,.
\end{eqnarray}
Twist-$4$ distribution amplitudes are:
\begin{eqnarray}
S_1(x_i)&=&6x_3(1-x_3)(\xi_4^0+\xi_4^{'0})\,,\hspace{1.5cm}P_1(x_i)=6(1-x_3)(\xi_4^0-\xi_4^{'0})\,,\nonumber\\
V_2(x_i)&=&0\,,\hspace{5.0cm}A_2(x_i)=-24x_1x_2\phi_4^0\,,\nonumber\\
V_3(x_i)&=&12(x_1-x_2)x_3\psi_4^0\,,\hspace{2.3cm}A_3(x_i)=-12x_3(1-x_3)\psi_4^0\,,\nonumber\\
T_2(x_i)&=&0\,,\hspace{5.0cm}T_3(x_i)=6(x_2-x_1)x_3(-\xi_4^0+\xi_4^{'0})\,,\nonumber\\
T_7(x_i)&=&-6(x_1-x_2)x_3(\xi_4^0+\xi_4^{'0})\,.
\end{eqnarray}
Twist-$5$ distribution amplitudes are:
\begin{eqnarray}
S_2(x_i)&=&\frac32(x_1+x_2)(\xi_5^0+\xi_5^{'0})\,,\hspace{1.5cm}P_2(x_i)=\frac32(x_1+x_2)(\xi_5^0-\xi_5^{'0})\,,\nonumber\\
V_4(x_i)&=&3(x_2-x_1)\psi_5^0\,,\hspace{2.8cm}A_4(x_i)=-3(1-x_3)\psi_5^0\,,\nonumber\\
V_5(x_i)&=&0\,,\hspace{4.9cm}A_5(x_i)=-6x_3\phi_5^0\,,\nonumber\\
T_4(x_i)&=&-\frac32(x_1-x_2)(\xi_5^0+\xi_5^{'0})\,,\hspace{1.3cm}T_5(x_i)=0\,,\nonumber\\
T_8(x_i)&=&-\frac32(x_1-x_2)(\xi_5^0-\xi_5^{'0})\,.
\end{eqnarray}
Finally twist-$6$ distribution amplitudes are:
\begin{eqnarray}
V_6(x_i)&=&0\,,\hspace{4.5cm}A_6(x_i)=-2\phi_6^0\,,\nonumber\\
T_6(x_i)&=&0\,.
\end{eqnarray}

\newpage

\begin{center}
\begin{minipage}{12cm}
\baselineskip 24pt {\sf Fig. 1.}{\quad Dependence of the four parameters $f_{\Sigma}$, $\lambda_1$, $\lambda_2$ and $\lambda_3$ of $\Sigma$ on the Borel parameter
$M_B^2$. The lines correspond to the threshold $s_0=2.65-2.85\; \mbox{GeV}^{2}$ from the bottom up.}
\end{minipage}
\end{center}

\begin{center}
\begin{minipage}{12cm}
\baselineskip 24pt {\sf Fig. 2.}{\quad Sum rules of the relative signs of the parameters on the Borel parameter. The threshold is used as $s_0=1.66^2\;
\mbox{GeV}^{2}$.}
\end{minipage}
\end{center}

\begin{center}
\begin{minipage}{12cm}
\baselineskip 24pt {\sf Fig. 3.}{\quad Dependence of the four parameters $f_\Lambda$, $\lambda_1$, $\lambda_2$ and $f_\Lambda\lambda_1^*$ of $\Lambda$ on the Borel
parameter $M_B^2$. The lines correspond to the threshold $s_0=2.45-2.65\; \mbox{GeV}^{2}$ from the bottom up.}
\end{minipage}
\end{center}

\begin{center}
\begin{minipage}{12cm}
\baselineskip 24pt {\sf Fig. 4.}{\quad Dependence of the form factor $G_M(Q^2)$ of $\Sigma$ on the Borel parameter at different momentum transfer. The lines
correspond to the points $Q^2=1\,,2\,,3\,,5\,,6\;\ \mbox{GeV}^{2}$ from the up down (left $\Sigma^+$) and from the bottom up (right $\Sigma^-$), respectively.}
\end{minipage}
\end{center}

\begin{center}
\begin{minipage}{12cm}
\baselineskip 24pt {\sf Fig. 5.}{\quad Fittings of the form factor $G_M(Q^2)$ by the dipole formula $\mu_{\Sigma^+}/(1+Q^2/m_0^2)^2$. The dashed lines are the
fittings, and figures $(a)\,,(b)$ correspond to the threshold $s_0=2.65\,,2.85\; \mbox{GeV}^2$, respectively.}
\end{minipage}
\end{center}

\begin{center}
\begin{minipage}{12cm}
\baselineskip 24pt {\sf Fig. 6.}{\quad Fittings of the form factor $G_M(Q^2)$ by the dipole formula $\mu_{\Sigma^-}/(1+Q^2/m_0^2)^2$. The dashed lines are the
fittings, and figures $(a)\,,(b)$ correspond to the threshold $s_0=2.65\,,2.85\; \mbox{GeV}^2$, respectively.}
\end{minipage}
\end{center}

\begin{center}
\begin{minipage}{12cm}
\baselineskip 24pt {\sf Fig. 7.}{\quad The $Q^2$-dependence of the form factor $G_M/(\mu_\Sigma G_D)$. The lines correspond to the threshold
$s_0=2.65\,,2.75\,,2.85\; \mbox{GeV}^{2}$ from the bottom up. The left corresponds to $\Sigma^+$ and the right corresponds to $\Sigma^-$}.
\end{minipage}
\end{center}

\newpage

\begin{table}
\renewcommand{\arraystretch}{1.1}
\caption{Independent baryon distribution amplitudes that enter the expansion in Eqs. (\ref{def-chi1}) to (\ref{def-chi6}).}
\begin{center}
\begin{tabular}{|l|l|l|l|}
\hline& Lorentz-structure & Light-cone projection & nomenclature
\\ \hline
twist-3 &  $ \left(C\!\not\!{z}\right) \otimes \!\not\!{z} $ &
$u^+_\uparrow u^+_\downarrow s^+_\uparrow$ & $\Phi_3(x_i) =
\left[V_1-A_1\right](x_i)$ \\ \hline&  $ \left(C i \sigma_{\perp z}
\right) \otimes \gamma^\perp \!\not\!{z} $ & $u^+_\uparrow
u^+_\uparrow s^+_\downarrow$ & $T_1(x_i)$ \\  \hline twist-4 &  $
\left(C\!\not\!{z}\right) \otimes \!\not\!{p} $ & $u^+_\uparrow
u^+_\downarrow s^-_\uparrow$ & $\Phi_4(x_i) =
\left[V_2-A_2\right](x_i)$
\\ \hline
& $ \left(C\!\!\not\!{z}\gamma_\perp\!\!\not\!{p}\,\right)
      \otimes \gamma^\perp\!\!\not\!{z} $
& $ u^+_\uparrow u^-_\downarrow s^+_\downarrow$ & $\Psi_4(x_i) =
\left[V_3-A_3\right](x_i)$ \\ \hline & $ \left(C
\!\not\!{p}\!\not\!{z}\right)  \otimes \!\not\!{z}$ & $u^-_\uparrow
u^+_\uparrow  s^+_\uparrow$ & $\Xi_4(x_i) = \left[T_3-  T_7 + S_1 +
P_1\right](x_i)$ \\ \hline & $ \left(C \!\not\!{p}\!\not\!{z}\right)
\otimes \!\not\!{z}$ & $u^-_\downarrow u^+_\downarrow  s^+_\uparrow$
& $\Xi_4'(x_i) = \left[T_3 + T_7 + S_1 - P_1\right](x_i)$\\ \hline&
$ \left(C i \sigma_{\perp z} \right) \otimes \gamma^\perp
\!\not\!{p} $ & $u^+_\downarrow u^+_\downarrow s^-_\downarrow$ &
$T_2(x_i)$
\\ \hline
twist-5 &  $ \left(C\!\not\!{p}\right) \otimes \!\not\!{z} $ &
$u^-_\uparrow u^-_\downarrow s^+_\uparrow$ & $\Phi_5(x_i) =
\left[V_5-A_5\right](x_i)$
\\ \hline
& $ \left(C\!\!\not\!{p}\gamma_\perp\!\!\not\!{z}\,\right) \otimes
\gamma^\perp\!\!\not\!{p} $ & $ u^-_\uparrow u^+_\downarrow
s^-_\downarrow $ & $\Psi_5(x_i) = \left[V_4-A_4\right](x_i)$ \\
\hline & $ \left(C \!\not\!{z}\!\not\!{p}\right)  \otimes
\!\not\!{p}$ & $u^+_\uparrow u^-_\uparrow  s^-_\uparrow$ &
$\Xi_5'(x_i) = \left[-T_4- T_8 + S_2 + P_2\right](x_i)$ \\
\hline & $ \left(C \!\not\!{z}\!\not\!{p}\right)  \otimes
\!\not\!{p}$ & $u^+_\downarrow u^-_\downarrow  s^-_\uparrow$ &
$\Xi_5(x_i) = \left[S_2 - P_2-T_4+ T_8\right](x_i)$ \\ \hline&  $
\left(C i \sigma_{\perp p} \right) \otimes \gamma^\perp \!\not\!{z}
$ & $u^-_\downarrow u^-_\downarrow s^+_\downarrow$ & $T_5(x_i)$
\\ \hline
twist-6 &  $ \left(C\!\not\!{p}\right) \otimes \!\not\!{p} $ &
$u^-_\uparrow u^-_\downarrow s^-_\uparrow$ & $\Phi_6(x_i) =
\left[V_6-A_6\right](x_i)$ \\ \hline&  $ \left(C i \sigma_{\perp p}
\right) \otimes \gamma^\perp \!\not\!{p} $ & $u^-_\uparrow
u^-_\uparrow s^-_\downarrow$ & $T_6(x_i)$ \\ \hline
\end{tabular}
\end{center} \label{tabDA-def}
\end{table}

\clearpage

\newpage

\begin{table}[h]
\caption{Magnetic moments of the $\Sigma$ baryons from various
models}
\begin{center}
\begin{tabular}{|c|c|c|c|c|c|c|c|c|c|}
\hline
$\mu(\mu_N)$& PDG & SR(1) &SR(2) & QCDSA & $\chi$PT & SKRM & LCSR(1) & LCSR(2) & Ours\\
\hline $\mu_{\Sigma^+}$& $2.46$ & $2.52$ & $3.30$ & $2.48$ & $2.458$
& $2.41$ &
$2.2$ & $2.9$ & $3.13$ \\
\hline $\mu_{\Sigma^-}$ & $-1.16$ & $-1.13$ & $-1.38$ &
$-0.90$ & $-1.16$ & $-1.10$ & $-0.8$ & $-1.1$ & $-1.59$\\
\hline
\end{tabular}
\end{center}\label{moment}
\end{table}

\clearpage

\newpage

\clearpage

\newpage

\clearpage

\newpage

\clearpage

\newpage

\clearpage

\newpage

\clearpage

\newpage

\clearpage

\newpage

\begin{thebibliography}{99}
\bibitem{exclusive} G. P. Lepage and S. J. Brodsky, Phys.\ Rev.\
Lett. {\bf 43} (1979) 545, 1625(E); S. J. Brodsky, G. P. Lepage and
A. A. Zaidi, Phys.\ Rev.\ D {\bf 23} (1981) 1152; S. J. Brodsky and
G. P. Lepage, Phys.\ Rev.\ D {\bf 24} (1981) 2848.
\bibitem{exclusive2} A. V. Efremov and A. V. Radyushkin, Phys.\
Lett.\ B. {\bf 94} (1980) 245; G. P. Lepage and S. J. Brodsky,
Phys.\ Lett.\ B. {\bf 874} (1979) 359; G. P. Lepage and S. J.
Brodsky, Phys.\ Rev.\ D {\bf 22} (1980) 2157.
\bibitem{Braun2} V. M. Braun and I. E. Filyanov, Z.\ Phys.\ C {\bf 48}
(1990) 239; P. Ball, JHEP {\bf 9901} (1999) 010.
\bibitem{Balitsky}I. I. Balitsky and V. M. Braun, Nucl.\ Phys.\ B {\bf
311} (1989) 541.
\bibitem{mesondas}P. Ball, V. M. Braun, Y. Koike and K. Tanaka, Nucl.\ Phys.\ B {\bf 529}
(1998) 323; P. Ball and V. M. Braun, Nucl.\ Phys.\ B {\bf 543} (1999) 201; P. Ball and M. Boglione, Phys.\ Rev.\ D {\bf 68} (2003) 094006; P. Ball, V. M. Braun and
A. Lenz, JHEP {\bf 0605} (2006) 004; P. Ball, V. M. Braun and A. Lenz, arXiv: 0707.1201.
\bibitem{Agaev} S. S. Agaev, Phy.\ Rev.\ D {\bf 72} (2005) 074020.
\bibitem{Yang} K. C. Yang, Nucl.\ Phys.\ B {\bf 776} (2007) 187.
\bibitem{review} V. M. Braun, arXiv: hep-ph/0608231.
\bibitem{Chernyak} V. L. Chernyak, A. A. Ogloblin, and L. R.
Zhitnitsky, Z.\ Phys.\ C {\bf 42} (1989) 569; Sov. J. Nucl. Phys.
{\bf 48} (1988) 536.
\bibitem{Braun1} V. M. Braun, R. J. Fries, N. Mahnke, and E. Stein, Nucl.\ Phys.\ B {\bf
589} (2000) 381.
\bibitem{Wang} M.Q. Huang and D.W. Wang, arXiv:hep-ph/0608170.
\bibitem{Ball}P. Ball, V. M. Braun and E. Gardi, arXiv: 0804.2424.
\bibitem{renom} V. M. Braun, A. N. Manashov and J. Rohrwild, arXiv:
0806.2531.
\bibitem{bdas} V. M. Braun, S. E. Derkachov, G. P. Korchemsky, and A.
N. Manashov, Nucl.\ Phys.\ B {\bf 553} (1999) 355.
\bibitem{ff} V. M. Braun, A. Lenz, N. Mahnke, and E. Stein, Phys.\
Rev.\ D {\bf 65} (2002) 074011; A. Lenz, M. Wittmann, and E. Stein, Phys.\ Lett.\ B {\bf 581} (2004) 199; V. M. Braun, A. Lenz, G. Peters, and A.V. Radyushkin,
Phys.\ Rev.\ D {\bf 73} (2006) 034020; V. M. Braun, A. Lenz, and M. Wittmann, Phys.\ Rev.\ D {\bf 73} (2006) 094019; V. M. Braun, D.Yu. Ivanov, A. Lenz, and A.
Peters, Phys.\ Rev.\ D {\bf 75} (2007) 014021.

\bibitem{Aliev} T. M. Aliev, K. Azizi, and A. Ozpineci, M. Savci,
arXiv: 0802.3008v2.
\bibitem{Wym} Y. M. Wang, Y. Li, and C. D. L\"{u}, arXiv: 0804.0648.
\bibitem{Wzg} Z. G. Wang, S. L. Wan, and W. M. Yang, Phys.\ Rev.\ D
{\bf 73} (2006) 094001; Z. G. Wang, S. L. Wan and W. M. Yang, Eur.\
Phys.\ J.\ C {\bf 47} (2006) 375.
\bibitem{Huang}
M.Q. Huang and D.W. Wang, phys.\ Rev.\ D {\bf 69} (2004) 094003.
\bibitem{SVZ} M. A. Shifman, A. I. Vainshtein and V. I. Zakharov, Nucl.\
Phys.\ B {\bf147} (1979) 385; B {\bf147} (1979) 448; V. A. Novikov,
M. A. Shifman, A. I. Vainshtein and V. I. Zakharov, Fortschr.\
Phys.\ {\bf 32} (1984) 11.
\bibitem{lcsr1} I. I. Balitsky, V. M. Braun
and A. V. Kolesnichenko, Nucl.\ Phy. B {\bf 312} (1989) 509; Sov.\
J.\ Nucl.\ Phys.\ {\bf 44} (1986) 1028; ibid.\ {\bf 48} (1988) 546.
\bibitem{lcsr2} V. M. Braun and I. E. Filyanov, Z.\ Phys.\ C {\bf
44} (1989) 157.
\bibitem{lcsr3} V. L. Chernyak and I. R. Zhitnitskii, Nucl.\ Phys.\
B {\bf 345} (1990) 137.

\bibitem{Bebek} C. J. Bebek $et$ $al.$, Phys.\ Rev.\ D {\bf 9} (1974) 1229; C. J. Bebek $et$ $al.$, Phys.\ Rev.\ D {\bf 13} (1976) 25;
C. J. Bebek $et$ $al.$, Phys.\ Rev.\ D {\bf 17} (1978) 1693.
\bibitem{Dally}E. B. Dally $et$ $al.$, Phys.\ Rev.\ Lett. {\bf 39} (1977) 1176; W. R. Molzon $et$ $al.$, Phys.\ Rev.\ Lett. {\bf 41} (1978)
1213 [Erratum-ibid. 41 (1978 ERRAT, 41,1835.1978) 1523]; E. B. Dally $et$ $al.$, Phys.\ Rev.\ Lett. {\bf 45} (1980) 232; E. B. Dally $et$ $al.$, Phys. Rev.\ Lett.\
{\bf 48} (1982) 375;
\bibitem{Liesenfeld}A. Liesenfeld $et$ $al.$ [A1 Collaboration], Phys.\ Lett.\ B {\bf 468} (1999) 20.
\bibitem{Volmer}J. Volmer $et$ $al.$ [The Jefferson Lab F(pi) Collaboration], Phys.\ Rev.\ Lett. {\bf 86} (2001) 1713.
\bibitem{Horn}T. Horn $et$ $al.$ [Fpi2 Collaboration], Phys.\ Rev.\ Lett. {\bf 97} (2006) 192001.
\bibitem{Tadevosyan}V. Tadevosyan $et$ $al.$ [Jefferson Lab F(pi) Collaboration], Phys.\ Rev.\ C {\bf 75} (2007) 055205.

\bibitem{Walker} R. C. Walker $et$ $al.$, Phys.\ Rev.\ D {\bf 49} (1994) 5671.
\bibitem{Andivahis} L. Andivahis $et$ $al.$, Phys.\ Rev.\ D {\bf 50} (1994) 5491.
\bibitem{Arrington} J. Arrington, Phys.\ Rev.\ C {\bf 68} (2003) 034325.
\bibitem{Christy} M. E. Christy $et$ $al.$ (E94110 Collaboration), Phys.\
Rev.\ C {\bf 70} (2004) 015206.
\bibitem{Bosted} P. E. Bosted $et.$ $al.$, Phys.\ Rev.\ Lett. {\bf 29}
(1992) 3841.
\bibitem{Lung} A. Lung $et$ $al.$, Phys. Rev.\ Lett.\ {\bf 70} (1993) 718.
\bibitem{Qattan} I. A. Qattan $et$ $al.$, Phys.\ Rev.\ Lett. {\bf 94} (2005)
142301.
\bibitem{Bourgeois} P. Bourgeois $et$ $al.$, Phys.\ Rev.\ Lett. {\bf 24} (2006)
212001.
\bibitem{Anklin} H. Anklin $et$ $al.$, Phys.\ Lett.\ B {\bf 428} (1998) 248.
\bibitem{Kubon} G. Kubon $et$ $al.$, Phys.\ Lett.\ B {\bf 524} (2002) 26.

\bibitem{Kim} H. C. Kim, A. Blotz, M. V. Polyakov, K. Goeke, Phys.\
Rev.\ D {\bf 53}, 4013 (1996)
\bibitem{Kubis}B. Kubis, T. R. Hemmert and U. G. Meissner, Phys.\ Lett.\  B {\bf 456}, 240
(1999); B. Kubis and U. G. Meissner, Eur.\ Phys.\ J.\  C {\bf 18}, 747 (2001).
\bibitem{Cauteren}T. Van Cauteren $et$ $al.$, Eur.\ Phys.\ J.\ A {\bf 20},
283 (2004); T. Van Cauteren $et$ $al.$, nucl-th/0407017.

\bibitem{LambdaEM} Y. L. Liu and M. Q. Huang, arXiv:0810.4973.
\bibitem{PDG}  C. Amsler $et$ $al.$ (Particle Data Group), Phys.\ Lett.\ B {\bf 667} (2008) 1.
\bibitem{Chiu} C. B. Chiu, J. Pasupathy and S. L. Wilson, Phys.\ Rev.\ D {\bf
33} (1986) 1961.
\bibitem{Kerbikov} B. O. Kerbikov and Yu. A. Simonov, Phys.\ Rev.\ D {\bf 62} (2000) 093016.
\bibitem{Puglia} S. J. Puglia and M. J. Ramsey-Musolf, Phys.\ Rev.\ D {\bf 62} (2000) 034010.
\bibitem{Park} N. W. Park and H. Weigel, Nucl.\ Phys.\ A {\bf 541} (1992) 453.
\bibitem{Aliev2} T. M. Aliev, A. Ozpineci and M. Savci. Phys.\ Rev.\ D {\bf 66} (2002) 016002, Erratum-ibid. D {\bf67} (2003)
039901.
\end{thebibliography}
\end{document}